\documentclass{emulateapj}

\usepackage{graphics} 
\usepackage{epsfig}
\usepackage{natbib}
\begin{document}

\title{Direct Measurement of Interstellar Extinction Toward Young Stars Using Atomic Hydrogen Lyman-$\alpha$ Absorption}

\author{Matthew McJunkin\altaffilmark{1}, Kevin France\altaffilmark{1,7}, P. C. Schneider\altaffilmark{2}, Gregory J. Herczeg\altaffilmark{3}, Alexander Brown\altaffilmark{1}, Lynne Hillenbrand\altaffilmark{4}, Eric Schindhelm\altaffilmark{5}, Suzan Edwards\altaffilmark{6}}

\altaffiltext{1}{Center for Astrophysics and Space Astronomy, University of Colorado, 389 UCB,
                        Boulder, CO 80309, USA; matthew.mcjunkin@colorado.edu}
\altaffiltext{2}{Hamburger Sternwarte, Gojenbergsweg 112, 21029 Hamburg, Germany} 
\altaffiltext{3}{Kavli Institute for Astronomy and Astrophysics, Peking University, Beijing 100871, China}  
\altaffiltext{4}{California Institute of Technology, Department of Astrophysics, MC105-24, 1200 E. California Blvd., Pasadena, CA 91125, USA} 
\altaffiltext{5}{Southwest Research Institute, 1050 Walnut Street, Suite 300, Boulder, CO 80302, USA}            
\altaffiltext{6}{Five College Astronomy Dept., Smith College, Northampton, MA 01063, USA}         
\altaffiltext{7}{NASA Nancy Grace Roman Fellow}   
  

\begin{abstract}

	Interstellar reddening corrections are necessary to reconstruct the intrinsic spectral energy distributions (SEDs) of accreting protostellar systems.  The stellar SED determines the heating and chemical processes that can occur in circumstellar disks.  Measurement of neutral hydrogen absorption against broad Lyman-$\alpha$ emission profiles in young stars can be used to obtain the total \ion{H}{1} column density (N(\ion{H}{1})) along the line of sight.  We measure N(\ion{H}{1}) with new and archival ultraviolet observations from the \textit{Hubble Space Telescope} ($HST$) of 31 classical T Tauri and Herbig Ae/Be stars.  The \ion{H}{1} column densities range from log$_{10}$(N(\ion{H}{1})) $\approx 19.6 - 21.1$, with corresponding visual extinctions of A$_{V}$ $= 0.02 - 0.72$ mag, assuming an R$_{V}$ of 3.1.  We find that the majority of the \ion{H}{1} absorption along the line of sight likely comes from interstellar rather than circumstellar material.  Extinctions derived from new $HST$ blue-optical spectral analyses,  previous IR and optical measurements, and new X-ray column densities on average overestimate the interstellar extinction toward young stars compared to the N(\ion{H}{1}) values by $\sim 0.6$ mag.  We discuss possible explanations for this discrepancy in the context of a protoplanetary disk geometry.              

\end{abstract}

\clearpage

\section{Introduction}

	Extinction correction of an observed short wavelength spectral energy distribution (SED) allows determination of the level of irradiance in the planet-forming environment of the protoplanetary disk (e.g. \citealt{2012A&amp;A...547A..69A}).  Ultraviolet (UV) radiation plays a particularly important role in gas heating (e.g. \citealt{2004A&amp;A...428..511J, 2007ApJ...661..334N, 2009A&amp;A...501..383W}) and protoplanetary disk gas chemistry (e.g. \citealt{1999A&amp;A...351..233A, 2009Sci...326.1675B, 2011ApJ...726...29F, 2013A&amp;A...559A..46B}).  Neutral hydrogen (\ion{H}{1}) Lyman-$\alpha$ in emission is a significant component of the far-UV radiation \citep{2003ApJ...591L.159B, 2004ApJ...607..369H, 2012ApJ...756L..23S} produced in the stellar atmosphere and accretion funnel flow (see, e.g., \citealt{1994ApJ...426..669H, 1998AJ....116..455M, 2006MNRAS.370..580K, 2007prpl.conf..479B, 2011ApJ...743..105I}).  

	Line of sight extinction values are typically estimated for young stars from measured color excess relative to an assumed intrinsic color corresponding to the spectral type.  However, due to veiling of the photosphere in Classical T Tauri Stars (CTTSs) by excess accretion emission (e.g. \citealt{1998ApJ...492..323G}), the extinction-free color determination becomes challenging without an accurate measurement of the excess flux due to accretion shocks \citep{1995ApJ...452..736H}.  For this reason, more sophisticated extinction estimates come from modeling of spectrophotometry, including contributions not only from the underlying stellar photosphere but also from a veiling continuum (e.g. \citealt{2003ApJ...583..334H, HerczegHill2013}).  
	
	Optical extinction measurements, which are often calculated using assumed colors for a particular spectral type (see Section 3.3), can vary due to uncertainties in the spectral type for the target star.  Scatter in intrinsic spectral color and effective temperature at a given spectral type ($\sim \pm 2000$ K at A0; \citealt{1999A&amp;AS..137..273G}) and rapid rotation which leads to large temperature gradients on the star \citep{2011A&amp;A...530A..85M} can also contribute to a broad range of optical extinction.  A further complication in deriving extinction from colors or spectrophotometry is that the optical variability of CTTSs can also be substantial ($\sim 0.1 - 2.5$ mag in the V band; \citealt{1994AJ....108.1906H}), which includes large color variability in many cases, and thus further uncertainty in the extinction estimates if non-simultaneous data is used.  The intrinsic color of a star may be inaccurately determined if the star is in a close binary system; however, this is not applicable to most of our targets (except possibly the spectroscopic binaries V4046 Sgr, TWA 3A, AK Sco, and HD 104237).  The X-ray and Lyman-$\alpha$ \ion{H}{1} column densities of these binaries should also not be strongly affected.       
	
\begin{figure} \figurenum{1}
\begin{center}
\includegraphics[angle=90,scale=0.35]{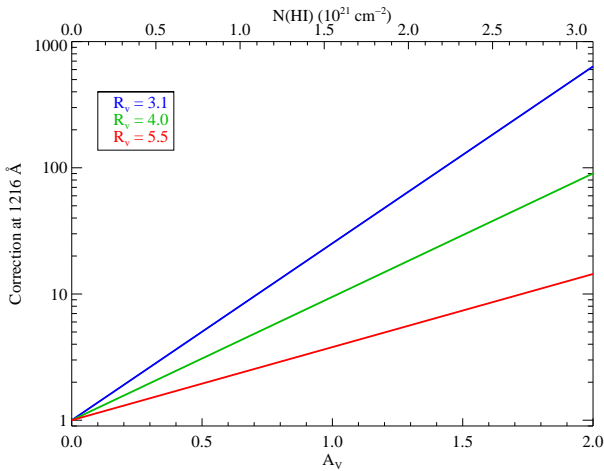}
\caption{
Correction to the normalized unreddened flux at 1216~\AA~ as a result of changing visual extinction, A$_{V}$, using the reddening curve of \citet{1989ApJ...345..245C} including the near-UV update of \citet{1994ApJ...422..158O}.  A$_{V}$ values are converted to N(\ion{H}{1}) values along the top of the figure using the \citet{1978ApJ...224..132B} relation assuming R$_{V} = 3.1$.  As the visual extinction increases from 0 to 2, the unreddened flux changes by up to factors of 600.  
}\label{fig:FUV_Correction}
\end{center}
\end{figure}

	Because of the range in techniques employed and wavelengths used, as well as the astrophysical uncertainties due to variability, A$_{V}$ measurements for the same star can range dramatically in the literature.  The reported visual extinction of HD 135344B spans the range 0.30 \citep{2006A&amp;A...459..837G} to 0.96 \citep{2013MNRAS.429.1001A}, and the visual extinction of RW Aur A has been cited from 0.50 \citep{2001ApJ...556..265W} to 1.58 \citep{2004ApJ...616..998W}.
Even small changes in the adopted extinction create large changes in estimates for the UV luminosity.  At an A$_{V}$ value of 2 mag, the normalized flux correction at Lyman-$\alpha$ (1216~\AA) is $\sim$ 600, while at A$_{V}$ = 0.5 mag, the correction is only $\sim 5$ (see Figure~\ref{fig:FUV_Correction}).  The uncertainty in the stellar UV radiation field due to extinction uncertainty has a significant effect on chemical models of disks.  
  
	The \ion{H}{1} column density is well correlated with interstellar reddening caused by dust grains through the well-known \citet{1978ApJ...224..132B} relation ($\langle$N(\ion{H}{1})/$E(B - V) \rangle = 4.8 \times 10^{21}$ atoms cm$^{-2}$ mag $^{-1}$).  This relation was derived from a correlation (typical scatter $\sim$ 50\% about the fit line for stars with accurate $E(B - V)$) of the interstellar \ion{H}{1} column densities and color excess ($E(B-V)$) values of 100 stars surveyed with the \textit{Copernicus} satellite.  The interstellar \ion{H}{1} columns were calculated by fitting the absorption from interstellar gas against the continuum emission of the stars.  \citet{1994ApJ...427..274D} found a similar correlation between \ion{H}{1} and $E(B-V)$ ($\langle$N(\ion{H}{1})/$E(B - V) \rangle = 4.93 \times 10^{21}$ atoms cm$^{-2}$ mag $^{-1}$) with 6\% uncertainty using archival Lyman-$\alpha$ absorption line data of 393 stars from the International Ultraviolet Explorer ($IUE$) satellite.
	
	In this work, we present new measurements of the \ion{H}{1} column density along the line of sight toward accreting young stars (spectral types roughly A0 to M4) using a UV-based technique.  We employ a least-squares fitting routine to simultaneously parameterize the stellar plus accretion Lyman-$\alpha$ emission profile and the line of sight absorption of \ion{H}{1}.  \citet{2003AJ....126.3076W}, \citet{2004ApJ...607..369H}, and \citet{2006AstL...32..176L} performed similar fitting routines to a smaller subset of CTTSs.  We fit the emission profile with a broad and narrow Gaussian component and employ a Voigt profile to characterize the \ion{H}{1} absorption.  Adopting the well-characterized relationship between the total hydrogen column and $E(B-V)$ \citep{1978ApJ...224..132B}, modulo assumptions about the grain distribution (R$_{V}$), we can make a straightforward determination of the reddening that is not confused by veiling or interpretation of (spectro)photometric variability.  
	
	We describe the targets and observations in Section 2.  The analysis of the observations (including new optical and X-ray measurements) and a description of the Lyman-$\alpha$ model fitting procedure is presented in Section 3.  The results of the fits and comparisons to results using different techniques in other wavelengths are described in Section 4.  We provide a discussion and offer possible explanations for the discrepancy in extinction values from this work compared to those measured in the optical, IR, and X-ray in Section 5.  Finally, Section 6 contains a summary of our results.

\section{Targets and Observations}

	We analyze Lyman-$\alpha$ spectra of 31 young stars listed in Table 1.  Many of these observations have been described previously in the molecular hydrogen (H$_{2}$) and hot gas surveys of \citet{2012ApJ...756..171F} and \citet{2013arXiv1304.3746A}, respectively. Several targets are binaries or multiples (DF Tau: \citealt{1993AJ....106.2005G}, DK Tau, HN Tau, RW Aur, and UX Tau: all \citealt{2006A&amp;A...459..909C}), but  we observe only the primary within the aperture.  TWA 3A, AK Sco, and HD 104237 are spectroscopic binaries \citep{2000ApJ...535L..47M, 2009ApJ...698L.108G, 2004A&amp;A...427..907B}.  V4046 Sgr is a known short-period binary system \citep{2000IAUS..200P..28Q} which nevertheless acts as a point source for the interstellar absorption.  The stars are young; ranging in age from Myr old members of star forming regions (Taurus-Auriga, Chamaeleon I, and $\eta$ Chamaeleontis) to $10 - 30$ Myr field pre-main sequence stars (e.g. TW Hya, AK Sco) with most targets being in the range of 1 - 10 Myr, comparable to the timescale of depletion for accreting gas and circumstellar dust (\citealt{2007ApJ...671.1784H, 2010A&amp;A...510A..72F}) and therefore presumably giant planet formation (\citealt{2005Icar..179..415H}).  The majority of the targets are located in the Taurus-Auriga, Chamaeleon I, and $\eta$ Chamaeleontis star forming regions.  The remaining targets belong to other associations and isolated systems.  The distances to all the targets are listed in Table 1.  

\begin{deluxetable*}{ccccccc}\tablenum{1}
\tabletypesize{\scriptsize}
\tablecaption{Target Parameters \label{tab:Target_Parameters}}
\tablewidth{0pt}
\tablehead{
\colhead{Object }& \colhead{Spectral} & \colhead{Distance}  & \colhead{L$_{*}$}& \colhead{M$_{*}$} & \colhead{\.M} & \colhead{Ref.\tablenotemark{b}}\\
&{Type} & (pc) &{(L$_{\odot}$)}& {(M$_{\odot}$)}&{($10^{-8}$~M$_{\odot}$ yr$^{-1}$)}& }

\startdata
AA Tau & K7 & 140 & 0.71 & 0.80 & 0.33 &  2,4,7,12,16,58,59\\
AB Aur & A0 & 140 & 46.8 & 2.4 & 1.8 & 19,49,50,58,59 \\
AK Sco & F5 & 103 & 7.59 & 1.35 & 0.09 & 18,20,34,62 \\
BP Tau & K7 & 140 & 0.925 & 0.73 & 2.88 & 7,12,38,58,59 \\
CS Cha & K6 & 160 & 1.32 & 1.05 & 1.20 & 21,35,60 \\
CV Cha & G8 & 160 & 7.7 & 2.00 & 3.16 & 22,36,60 \\
DE Tau & M0 & 140 & 0.87 & 0.59 & 2.64& 7,10,12,58,59\\
DF Tau A & M2 & 140 & 1.97 & 0.19 & 17.7 & 7,10,58,59 \\
DK Tau A & K7 & 140 & 1.45 & 0.71 & 3.79 & 7,10,12,58,59 \\
DM Tau & M1.5 & 140 & 0.24 & 0.50 & 0.29 & 16,29,32,58,59 \\
DN Tau & M0 & 140 & 0.87 & 0.60 & 0.35 & 7,16,32,39,58,59 \\
DR Tau & K7 & 140 & 1.09 & 0.80 & 3.16 &  2,8,16,58,59 \\
GM Aur & K5.5 & 140 & 0.74 & 1.20 & 0.96 & 7,16,32,58,59 \\
HD 100546  & B9.5 & 103 & 32.4 & 2.4 & 0.1 & 19,52,53,65 \\
HD 104237  & A7.5 & 116 & 34.7 & 2.50 & 3.50 & 19,23,31,45 \\
HD 135344B  & F3 & 140 & 8.13 & 1.60 & 0.54 & 19,31,42,64 \\
HD 163296 & A1 & 122 & 24.0 & 2.0 & 6.9 & 18,19,50,51\\
HN Tau A & K5 & 140 & 0.19 & 0.85 & 0.13 & 6,7,12,58,59\\
IP Tau & M0 & 140 & 0.41 & 0.68 & 0.08 & 7,12,58,59 \\
LkCa 15 & K3 & 140 & 0.72 & 0.85 & 0.13 & 12,29,32,58,59 \\
RECX-11 & K4 & 97 & 0.59 & 0.80 & 0.03 & 13,24,47,61 \\
RECX-15 & M2 & 97 & 0.08 & 0.40 & 0.10& 13,14,15,61\\
RU Lup & K7 & 121 & 0.42 & 0.80 & 3.00 & 25,30,41,62 \\
RW Aur A & K4 & 140 & 2.3 & 1.40 & 3.16 & 5,9,11,12,17,58,59\\
SU Aur & G1 & 140 & 9.6 & 2.30 & 0.45 & 1,3,8,11,12,58,59\\
SZ 102 & K0 & 200 & 0.01 & 0.75 & 0.08 & 26,37,43,48 \\
TWA 3A & M4Ve & 34 & 0.09 & 0.15 & 0.005 & 54,55,56,57,63 \\
TW Hya & K6 & 54 & 0.17 & 0.60 & 0.02 & 27,30,42,62 \\
UX Tau A & K2 & 140 & 3.5 & 1.30 & 1.00 & 12,32,58,59 \\
V4046 Sgr & K5 & 83 & 0.5+0.3 & 0.86+0.69 & 1.30 & 28,33,44 \\
V836 Tau & K7 & 140 & 0.32 & 0.75 & 0.01 & 12,30,46,58,59 \\

\enddata
\tablenotetext{a}{~ (1) \citet{2002ApJ...566.1124A}; (2) \citet{2007ApJ...659..705A}; (3) \citet{1988ApJ...330..350B}; (4) \citet{1999A&amp;A...349..619B}; (5) \citet{2007ApJ...669.1072E}; (6) \citet{2011ApJ...734...31F}; (7) \citet{1998ApJ...492..323G}; (8) \citet{2000ApJ...544..927G}; (9) \citet{1995ApJ...452..736H}; (10) \citet{2001ApJ...561.1060J}; (11) \citet{2000ApJ...539..815J}; (12) \citet{2009ApJ...704..531K}; (13) \citet{2004MNRAS.351L..39L}; (14) \citet{2004ApJ...609..917L}; (15) \citet{2007MNRAS.379.1658R}; (16) \citet{2010A&amp;A...512A..15R}; (17) \citet{2001ApJ...556..265W}; (18) \citet{1998A&amp;A...330..145V}; (19) \citet{2005A&amp;A...437..189V}; (20) \citet{2003A&amp;A...409.1037A}; (21) \cite{1996MNRAS.280.1071L}; (22) \citet{2000A&amp;A...358..593S}; (23) \citet{2003ApJ...599.1207F}; (24) \citet{2001MNRAS.321...57L}; (25) \citet{2005AJ....129.2777H}; (26) \citet{2010A&amp;A...511A..10C}; (27) \citet{1999ApJ...512L..63W}; (28) \citet{2000IAUS..200P..28Q}; (29) \citet{1998ApJ...495..385H}; (30) \citet{2008ApJ...681..594H}; (31) \citet{2006A&amp;A...459..837G}; (32) \citet{2011ApJ...732...42A}; (33) \citet{2011ApJ...729....7F}; (34) \citet{2009ApJ...698L.108G}; (35) \citet{2007ApJ...664L.111E}; (36) \citet{2009MNRAS.398..189H}; (37) \citet{2003A&amp;A...406.1001C}; (38) \citet{2000ApJ...545.1034S}; (39) \citet{2003ApJ...597L.149M}; (40) \citet{2010ApJ...717..441E}; (41) \citet{2007A&amp;A...461..253S}; (42) \citet{2008ApJ...684.1323P}; (43) \citet{2004ApJ...604..758C}; (44) \citet{2010ApJ...720.1684R}; (45) \citet{2004ApJ...608..809G}; (46) \citet{2008ApJ...687.1168N}; (47) \citet{2011ApJ...743..105I}; (48) \citet{1994AJ....108.1071H}; (49) \citet{2011ApJ...729L..17H}; (50) \citet{2011AJ....141...46D}; (51) \citet{2007A&amp;A...469..213I}; (52) \citet{2007ApJ...665..512A}; (53) \citet{2005ApJ...620..470G}; (54) \citet{2008hsf2.book..757T}; (55) \citet{2004AJ....128.1812D}; (56) \citet{2009A&amp;A...508..833D}; (57) \citet{2000ApJ...535L..47M}; (58) \citet{1999A&amp;A...352..574B}; (59) \citet{2007ApJ...671..546L}; (60) \citet{2004ApJ...602..816L}; (61) \citet{1999PASA...16..257M}; (62) \citet{2007A&amp;A...474..653V}; (63) \citet{2005ApJ...634.1385M}; (64) \citet{2009ApJ...699.1822G}; (65) \citet{1997A&amp;A...324L..33V}.}

\end{deluxetable*}

	The sample data was obtained with the \emph{Hubble Space Telescope} Cosmic Origins Spectrograph ($HST$-COS) and Space Telescope Imaging Spectrograph (STIS).  

\subsection{COS Observations}

	Observations from the DAO of Tau guest observing program (PID 11616; PI - G. Herczeg) comprise the majority of the data.  Additional COS observations include those from the COS Guaranteed Time Observing program (PIDs 11533 and 12036; PI - J. Green) and observations of HD 135344B.  Most of the CTTS spectra were obtained using the far-UV medium-resolution modes of COS (G130M and G160M ($\Delta$$v \approx 18$ km s$^{-1}$ at Lyman-$\alpha$); \citealt{2012ApJ...744...60G}).  Multiple central wavelengths and several focal-plane positions covered the wavelength region from $\approx 1150 - 1750$ \AA~ while minimizing fixed pattern noise.  The far-UV COS data were processed using the COS calibration pipeline, CALCOS, and aligned and co-added with the procedure described in \citet{2010ApJ...720..976D}.  COS, a slitless spectrograph, experiences strong contamination from geocoronal Lyman-$\alpha$ filling the large (2.5" diameter) aperture.   We mask the central region of our Lyman-$\alpha$ spectra due to geocoronal \ion{H}{1} contamination.

\subsection{STIS Observations}	

	Targets exceeding the COS bright-object limit (AK Sco, CV Cha, and HD 104237) had to be observed with the E140M medium-resolution mode of STIS ($\Delta$$v \approx 7$ km s$^{-1}$ between 1150 and 1700\AA; \citealt{1998ApJ...492L..83K, 1998PASP..110.1183W}).  The observations were taken through the 0.2" $\times$ 0.2" slit for two to three orbits per object.  Archival STIS observations of RU Lupi, TW Hya, and the Herbig Ae/Be stars HD 100546, AB Aur, and HD 163296 along with observations of HD 104237 and TWA 3A with the G140M mode of STIS ($\Delta$$v \approx 30$ km s$^{-1}$ between 1150 and 1700\AA) complete the UV sample.  The STIS echelle calibration software developed for the StarCAT catalog (\citealt{2010ApJS..187..149A}, T. Ayres 2011, private communication) combined the far-UV STIS spectra.  Additionally, we use STIS G430L ($\Delta$$v \approx 600$ km s$^{-1}$ between 3050 and 5550\AA) blue optical spectra of CV Cha, HD 104237, RU Lupi, SU Aur, and AK Sco (PID 11616; PI - G. Herczeg) to calculate optical measurements of the visual extinction.  The geocoronal signal is weaker in the STIS data due to the narrower slit; however, we remove the inner region (typically 0.5 - 2 \AA) in all the spectra for consistency. 
		
\section{N(\ion{H}{1}) Analysis}

\subsection{Overview of Lyman-$\alpha$ Profiles }

	Lyman-$\alpha$ emission is the result of the electron in an \ion{H}{1} atom transitioning from the 2$p$ to the 1$s$ state.  The radiative lifetime of the 2$p$ level is $\sim 2 \times 10^{-9}$ s \citep{2009JPCRD..38..565W}, such that at interstellar densities the collisional depopulation of the state is negligible.  At large optical depths, the emitted Lyman-$\alpha$ photons will be reabsorbed and reemitted in different directions several times by other \ion{H}{1} atoms in the vicinity.  Due to the velocity of the \ion{H}{1} atoms, the scattered Lyman-$\alpha$ photons undergo a frequency shift (frequency scattering), which tends to move the photons away from the Lyman-$\alpha$ line center and create a Lyman-$\alpha$ profile with broad wings.  High infall velocities of \ion{H}{1} in the accretion flow are able to significantly broaden the Lyman-$\alpha$ profile (up to several hundred or even a thousand km s$^{-1}$).  Due to both spatial and spectral diffusion, the photons eventually escape the Lyman-$\alpha$ emitting region, and can then be absorbed by \ion{H}{1} along the line of sight of our observations.
	
	The most salient features of the Lyman-$\alpha$ profiles in young stars are their spectrally broad emission lines, typically extending out to several hundred km s$^{-1}$ on both the blue and red sides, and their strong central absorption.  The breadth of the emission line points to an accretion origin, such that this emission is likely produced close to the star (see Section 5.1).  Both the red side emission and blue side emission have contributions from accreting material.  The blue side emission is affected by outflowing material, while the central absorption is dominated by the damped interstellar component, though some contribution from self-absorption in the accretion or wind flow may also be present.  We ignore the effect of self-absorption in this analysis and do not expect the results or interpretations to be strongly influenced.     
	
\subsection{Lyman-$\alpha$ Profile Fitting}
	
\begin{figure} \figurenum{2}
\begin{center}
\includegraphics[angle=90,scale=0.35]{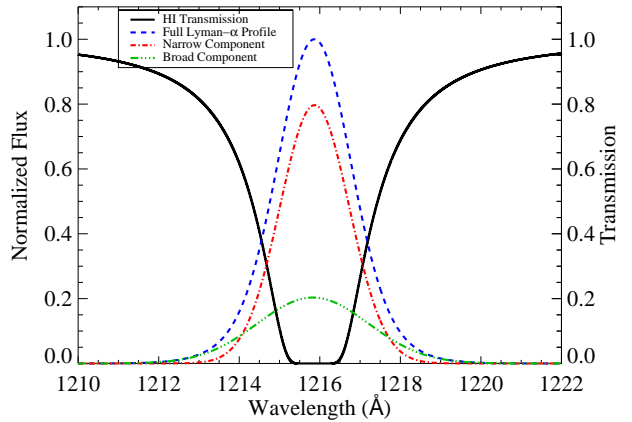}
\caption{
Components of the normalized Lyman-$\alpha$ model for RECX-15.  The broad component of the Lyman-$\alpha$ emission line is the green triple-dot-dashed line, the narrow component is the red dot-dashed line, and the unabsorbed model Lyman-$\alpha$ emission (equal to the broad component added to the narrow component) is marked by the blue dashed line.  The \ion{H}{1} transmission curve is overplotted in a solid black line.  
}\label{fig:Lya_components}
\end{center}
\end{figure}

\begin{figure} \figurenum{3}
\begin{center}
\includegraphics[angle=90,scale=0.35]{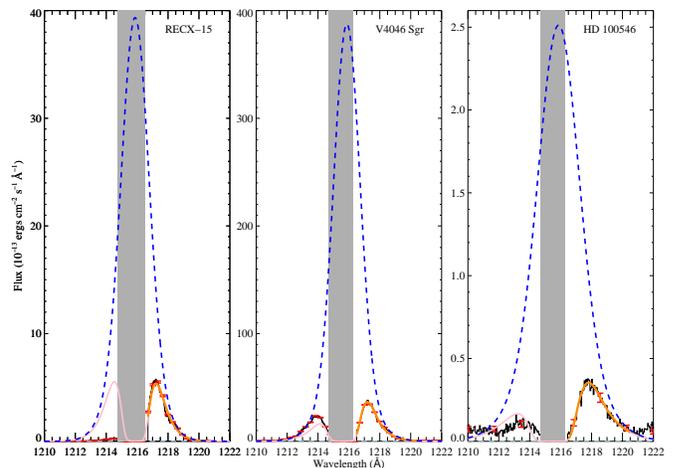}
\caption{
Model fits of the Lyman-$\alpha$ emission line in three representative targets (two CTTSs, one with and one without blueward emission, and a Herbig star).  The data (with the geocoronal Lyman-$\alpha$ emission masked out in the shaded region) is in black, with the absorbed profile (Lyman-$\alpha$ emission plus \ion{H}{1} absorption) in pink.  Only the redward side of the emission line is used in the least-squares fit due to stellar outflows affecting the blueward side, and this fit is highlighted in orange.  Selected error bars are shown in red.  The unabsorbed model Lyman-$\alpha$ emission is marked by the blue dashed line.    
}\label{fig:full_profiles}
\end{center}
\end{figure}

	We fit the Lyman-$\alpha$ spectra with a three component, nine parameter model consisting of broad and narrow Gaussian stellar emission lines and a Voigt \ion{H}{1} absorption profile to determine the best-fit interstellar \ion{H}{1} column density.  The models of the broad and narrow emission lines are each characterized by a heliocentric velocity, an amplitude, and a full-width half-maximum (FWHM) value, while the \ion{H}{1} absorption profile is characterized by a Doppler $b$-value, a heliocentric velocity, and a column density.  The broad and narrow components of the Lyman-$\alpha$ emission profile are pictured in Figure~\ref{fig:Lya_components} for RECX-15 to illustrate the model Lyman-$\alpha$ profile decomposition.  Fits showing the full Lyman-$\alpha$ model profiles for 3 select targets are shown in Figure~\ref{fig:full_profiles}.  These are not fully reconstructed profiles based on molecular fluorescence line fluxes such as those in \citet{2004ApJ...607..369H} and \citet{2012ApJ...746...97S, 2012ApJ...756L..23S}, and are shown only to illustrate the full fitting procedure.  Because many of the targets have outflows that absorb the blueward side of the Lyman-$\alpha$ emission line \citep{2012ApJ...756L..23S}, we restrict our fits to the redward side in order to derive an accurate determination of the interstellar \ion{H}{1} column density.  There is typically greater optical depth on the blueward side of the line, but without an outflow component in our model, we do not attempt to reproduce the full emission line.  We only fit a small region around the emission line such that a continuum parameter was not necessary in our model.  Adding a continuum model parameter to selected targets after the initial grid search gave no change to the best-fit column density.  Our restricted fitting region also allowed us to fit targets with an additional, very broad ($\Delta$$v \gtrsim \pm 1500$ km s$^{-1}$) Lyman-$\alpha$ emission component as we could isolate the narrower Lyman-$\alpha$ emissions and mask out the broad features when performing the fit.  These very broad profiles can be seen in BP Tau and GM Aur most prominently, and will be discussed in a future work.  

	We began with a preliminary fit by eye (which assumed an \ion{H}{1} velocity shift of 0 km s$^{-1}$) to determine initial parameter ranges and then performed a grid search, varying the parameters in order to find the best-fit model to the data.  Starting with a close fit to the data allowed us to use a higher resolution grid search and reduce the computational time needed to find the best fit.   

	We varied the model parameters in the grid search in order to find the best-fit values through a least-squares method similar to the procedure described in \citet{2013ApJ...766...12M}.  To decrease the computation time, the velocity shift of the broad and narrow emission line were set to a constant for all the targets (39 km s$^{-1}$ and 49 km s$^{-1}$ for the broad and narrow emission, respectively) based off the initial fit of V4046 Sgr (a high signal-to-noise prototypical example) to keep the fits consistent.  For the majority of the targets, the instrumental resolution was $\Delta$$v \approx 18$ km s$^{-1}$, so these offsets are only $\approx 2 - 2.5$ times the resolution element, and are mostly negligible compared to the breadth of both the Lyman-$\alpha$ emission and damped \ion{H}{1} absorption.  Assuming that neutral deuterium (\ion{D}{1}) traces \ion{H}{1} in the interstellar medium (ISM), we took the average of the \ion{D}{1} Doppler $b$-value from \citet{2004ApJ...602..776R} ($\sim 7.5$ km s$^{-1}$), and corrected for the D/H mass difference to give an \ion{H}{1} ISM Doppler width of $\sim 10$ km s$^{-1}$, and adopted this throughout.  The choice of $b$-value does not significantly affect the derived column densities at low $b$-values.  At the high \ion{H}{1} column densities of the initial model fits, we are well into the damping wing (``square root'') portion of the curve of growth, which has little dependence on Doppler $b$-value as long as $b \lesssim 100$ km s$^{-1}$.  We thus used three fixed parameters and six free parameters in our full Lyman-$\alpha$ absorption model.  
	
	Centered on our initial fit, we took eleven grid points in steps of 40 km s$^{-1}$ for both the broad and narrow emission FWHM, seven grid points in steps which were 30\% of the size of the initial fit values for the broad and narrow emission amplitudes, and eleven grid points in steps of 0.025 dex for the \ion{H}{1} absorption column density.  For the \ion{H}{1} absorption velocity shift, we chose thirteen grid points ranging from -70 km s$^{-1}$ to 50 km s$^{-1}$ in steps of 10 km s$^{-1}$.  This absorption velocity grid was chosen based on the observed range of stellar radial velocities, from $\sim -7$ km s$^{-1}$ (V4046 Sgr; \citealt{2006yCat.3249....0M}) to $+20$ km s$^{-1}$ (DR Tau; \citealt{2000AJ....119.1881A}), and the velocity of \ion{D}{1} in the ISM, which ranges from $-43$ to $+ 33$ km s$^{-1}$ for the local ($d < 100$ pc) Milky Way (\citealt{2004ApJ...602..776R}).            

	Due to the presence of protostellar outflows, the velocity of the absorbing gas is difficult to determine.  Changing the velocity shift of the \ion{H}{1} absorber relative to the emission lines affects the best-fit column density obtained from the model.  Thus, we performed two other grid searches to test the N(\ion{H}{1}) dependence on \ion{H}{1} velocity in the model: one where we allowed the \ion{H}{1} velocity to float around a velocity chosen from a new initial fit by eye (which did not assume an \ion{H}{1} velocity shift of 0 km s$^{-1}$), and one with a constant \ion{H}{1} velocity of 0 km s$^{-1}$ (which was typically close to the center of the absorption profile).  The floating velocity search had seven grid points in steps of 10 km s$^{-1}$ centered on the \ion{H}{1} velocity shift of the new initial fit to better estimate possible systematic errors in the column density due to the uncertainty in the \ion{H}{1} absorber velocity.  This grid search had the same grids as the first search in all other parameters.  Many of the model fits from this floating \ion{H}{1} velocity grid search had large \ion{H}{1} velocities which systematically shifted the absorption center to longer wavelengths, requiring smaller column densities derived from the redward side of the Lyman-$\alpha$ line to fit the data, which we take into account in our error budget.  The constant zero velocity search, however, had little effect on the best-fit column density from our first grid search.  The targets with negative \ion{H}{1} absorption velocities in the initial fit decreased their \ion{H}{1} column density (if they changed at all) in the zero velocity fit because of the shift to longer wavelengths similar to the floating velocity search.  The targets with positive \ion{H}{1} absorption velocities in the initial fit tended to increase their \ion{H}{1} column density.  The majority of the \ion{H}{1} column densities, however, did not change, and those that did only changed by $\pm 0.1-0.3$ dex.  This is because many of the best-fit \ion{H}{1} velocities from the first search were already close to zero, so that the velocity shift was minimal.       
	
\begin{figure*} \figurenum{4a}
\begin{center}
\plotone{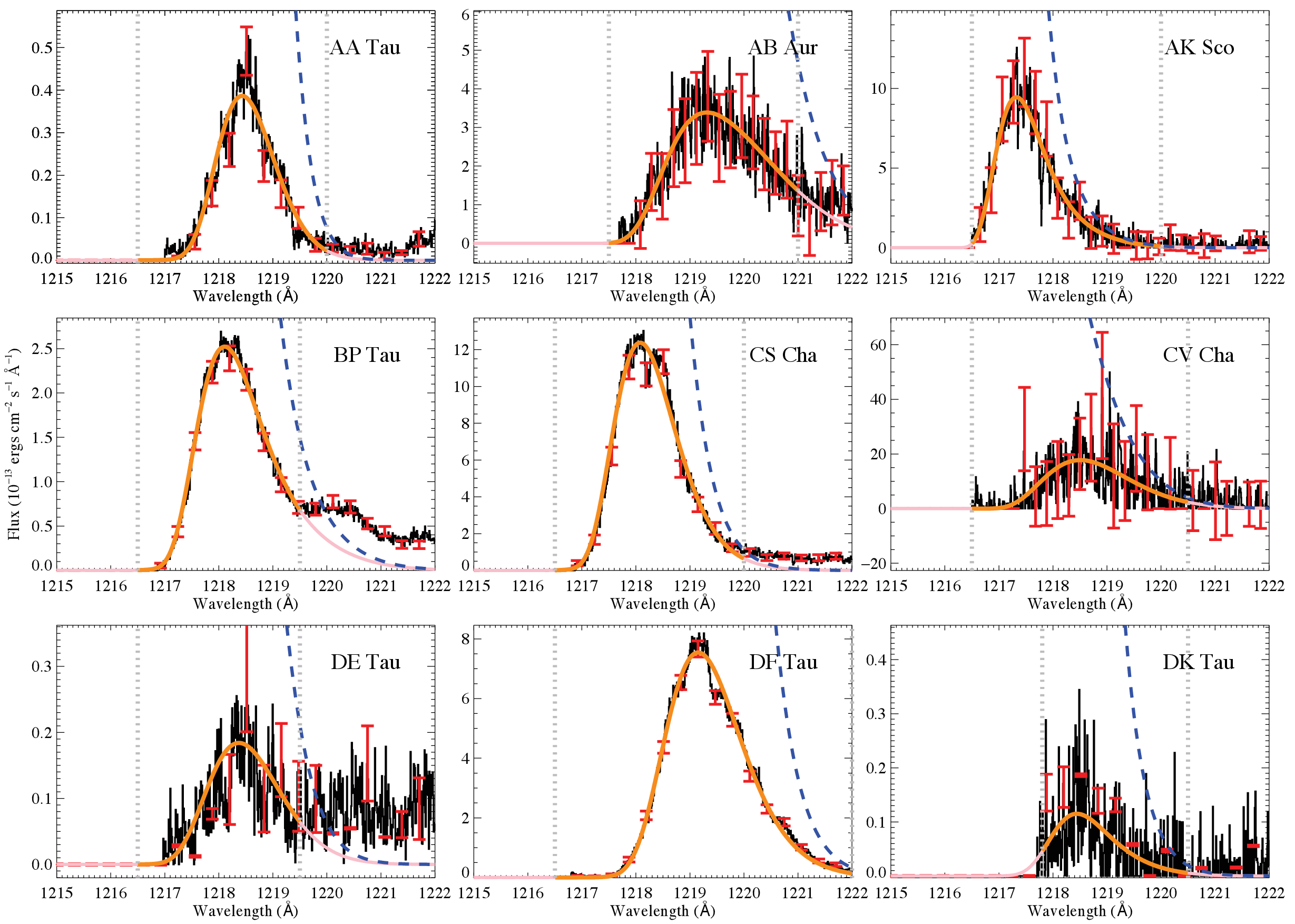}
\caption{
Model fits of the red side of the Lyman-$\alpha$ emission line.  The data, with geocoronal Lyman-$\alpha$ emission masked (zero flux is assumed in the region) is in black, with the fit in orange.  Selected error bars are shown in red and the wavelength region that is fit for each target is marked by horizontal dashed lines.  The unabsorbed model Lyman-$\alpha$ emission is marked by the blue dashed line.  Some targets have less reliable fits due to noise in the data and some fits may have failed to find a detectable redward Lyman-$\alpha$ emission line.  These targets (CV Cha, DE Tau, DK Tau, DM Tau, DN Tau, HN Tau, IP Tau, UX Tau, and V836 Tau) are identified with pink points in Figure~\ref{fig:N(H)vsAv}.  
}\label{fig:montage1}
\end{center}
\end{figure*}

\begin{figure*} \figurenum{4b}
\centering
\plotone{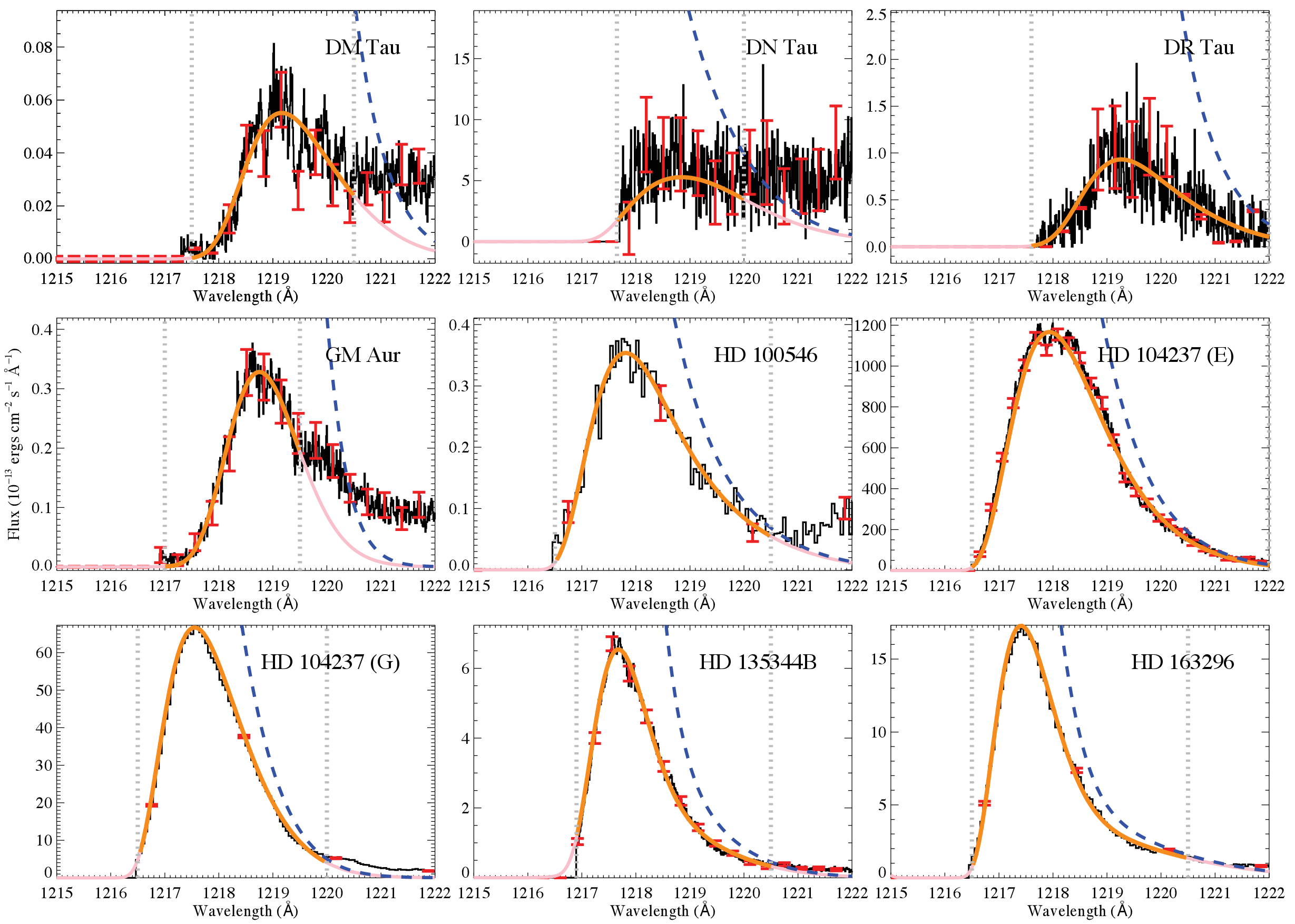}
\caption{
Same as 4a.
}\label{fig:montage2}
\end{figure*}

\begin{figure*} \figurenum{4c}
\begin{center}
\plotone{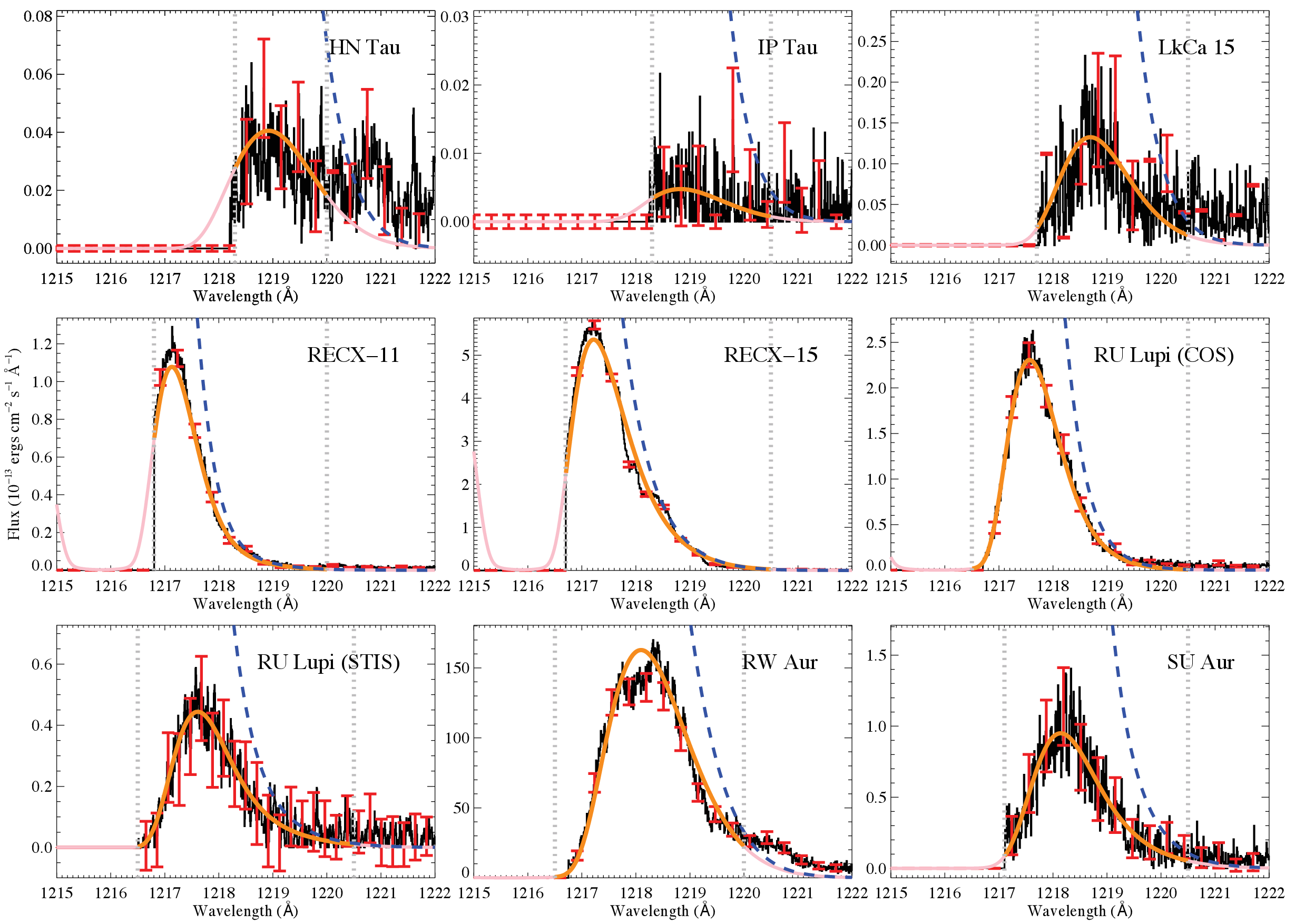}
\caption{
Same as 4a.
}\label{fig:montage3}
\end{center}
\end{figure*}

\begin{figure*} \figurenum{4d}
\begin{center}
\plotone{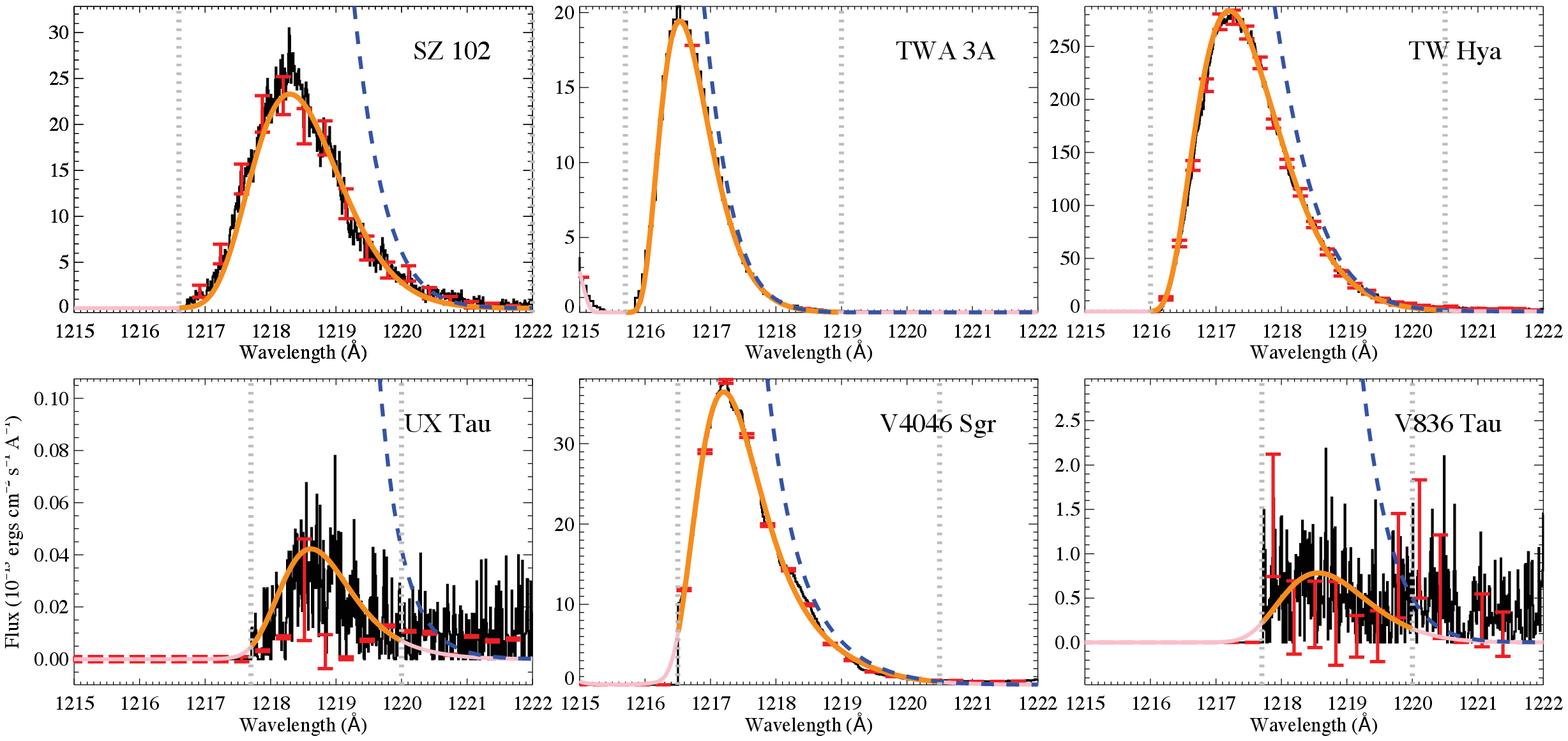}
\caption{
Same as 4a. 
}\label{fig:montage4}
\end{center}
\end{figure*}

	A montage of the model fits from our first grid search with constrained velocities for all of the targets can be seen in Figures~\ref{fig:montage1} - \ref{fig:montage4}.  The best-fit \ion{H}{1} column densities (from the first search) with errors are listed in Table 2.  The best-fit values of additional parameters that were allowed to vary in the grid search are listed in Table 3.  To estimate the errors, we set the model parameters to their best-fit values and varied the column density in steps of 0.025 dex.  Following the $\chi^{2}$ probability distribution for 1 degree of freedom, $\Delta \chi^{2} = 1$ defines a 68\% probability region.  We increased the minimum $\chi^{2}$ of our best-fit model by unity to obtain an estimate of the column density parameter range.  The initial errors were defined as the width of this range.  To account for the change in velocity systematically shifting the column densities lower (see above), we added the column density difference between the constrained and floating velocity grid searches in quadrature to the lower error bars on the column densities.  

\subsection{Extinction from Optical/Infrared Colors and Spectrophotometry}

	Previously for our sample, extinction has been estimated from a variety of techniques applied to optical or near-infrared data, the wavelengths at which the stellar photospheric flux is at a maximum.  These include extinction:
1) calculated by comparing the observed stellar colors to normal main-sequence stellar colors (e.g., \citealt{1995ApJS..101..117K}), 2) computed from $V$ and $(B-V)$ magnitudes (e.g., \citealt{1998A&amp;AS..133...81T}), 3) estimated by comparing the SED of the star to accretion emission of less veiled T Tauri stars (e.g., \citealt{2000ApJ...544..927G}), 4) derived from Two Micron All Sky Survey magnitudes with assumed $J-H$ or $J-K$ photospheric colors and a standard interstellar reddening law (e.g., \citealt{2004ApJ...616..998W, 2011ApJS..195....3F}), or 5) assessed from Paschen and Brackett line ratios compared to local line excitation models (e.g., \citealt{EdwardsInPrep}).  Recently, near-IR extinctions have been determined by fitting the observed target flux, the extinction corrected photospheric template flux, and the veiling of the photospheric template at a given wavelength \citep{2011ApJ...730...73F,2013ApJ...769...73M}.  

\begin{deluxetable*}{ccccccccc}\tablenum{2}
\tabletypesize{\scriptsize}
\tablecaption{Extinction Values \label{tab:newNHandAv}}
\tablehead{
\colhead{Target} & \colhead{$HST$ PID\tablenotemark{a}} & \colhead{PI}  & \colhead{A$_{V}$(lit)} & \colhead{A$_{V}$(lit) Ref.\tablenotemark{b,c}} &\colhead{log$_{10}$(N(H I))} & \colhead{A$_{V}$(3.1)\tablenotemark{d}} & \colhead{A$_{V}$(4.0)\tablenotemark{d}} & \colhead{A$_{V}$(5.5)\tablenotemark{d}} }

\startdata
\multicolumn{9}{c}{}\\
AA Tau & 11616 & G. Herczeg & 0.49 & 1 & $20.73^{+ 0.03}_{- 0.13}$ & 0.34 & 0.44 & 0.61 \\
AB Aur & $8065-S$ & B. Woodgate & 0.5 & 2 & $20.90^{+ 0.05}_{- 0.11}$ & 0.51 & 0.66 & 0.91 \\
AK Sco & $11616-S$ & G. Herczeg & 0.5 & 3 & $20.10^{+ 0.08}_{- 0.35}$ & 0.08 & 0.10 & 0.14 \\
BP Tau & 12036 & J. Green & 0.49 & 1 & $20.43^{+ 0.03}_{- 0.13}$ & 0.17 & 0.22 & 0.30 \\
CS Cha & 11616 & G. Herczeg & 0.8 & 4 & $20.400^{+ 0.03}_{- 0.15}$ & 0.16 & 0.21 & 0.29 \\
CV Cha & $11616-S$ & G. Herczeg & 1.67 & 5 & $20.45^{+ 0.05}_{- 0.16}$ & 0.18 & 0.23 & 0.32 \\
DE Tau & 11616 & G. Herczeg & 0.59 & 1 & $20.68^{+ 0.13}_{- 0.18}$ & 0.31 & 0.39 & 0.54 \\
DF Tau A & 11533 & J. Green & 0.60 & 6 & $20.93^{+ 0.03}_{- 0.03}$ & 0.54 & 0.70 & 0.96 \\
DK Tau A & 11616 & G. Herczeg & 0.76 & 1 & $20.85^{+ 0.23}_{- 0.05}$ & 0.46 & 0.59 & 0.81 \\
DM Tau & 11616 & G. Herczeg & 0.0 & 1 & $20.88^{+ 0.05}_{- 0.09}$ & 0.48 & 0.62 & 0.86 \\
DN Tau & 11616 & G. Herczeg & 1.89 & 7  & $20.45^{+ 0.13}_{- 0.14}$ & 0.18 & 0.23 & 0.32 \\
DR Tau & 11616 & G. Herczeg & 1.2 & 8 & $20.88^{+ 0.10}_{- 0.08}$ & 0.48 & 0.62 & 0.86 \\
GM Aur & 11616 & G. Herczeg & 0.14 & 1 & $20.90^{+ 0.03}_{- 0.35}$ & 0.51 & 0.66 & 0.91 \\
HD 100546 & $8895-S$ & S. Heap & 0.36 & 9 & $20.13^{+ 0.05}_{- 0.38}$ & 0.09 & 0.11 & 0.15 \\
HD 104237 (E140M) & $11616-S$ & G. Herczeg & 0.7 & 2 & $20.15^{+ 0.03}_{- 0.40}$ & 0.08 & 0.12 & 0.16 \\
HD 104237 (G140M) & $9241-S$ & A. Danks & 0.7 & 2 & $20.03^{+ 0.03}_{- 0.30}$ & 0.07 & 0.09 & 0.12 \\
HD 135344B & 11828 & A. Brown & 0.30 & 2 & $20.28^{+ 0.03}_{- 0.27}$ & 0.12 & 0.16 & 0.22 \\
HD 163296 & $8065-S$ & B. Woodgate & 0.30 & 2 & $19.90^{+ 0.03}_{- 0.40}$ & 0.05 & 0.07 & 0.09 \\
HN Tau A & 11616 & G. Herczeg & 0.52 &  1 & $20.75^{+ 0.13}_{- 0.05}$ & 0.36 & 0.47 & 0.64 \\
IP Tau & 11616 & G. Herczeg & 0.24 & 1 & $21.05^{+ 0.40}_{- 0.32}$ & 0.72 & 0.94 & 1.29 \\
LkCa 15 & 11616 & G. Herczeg & 0.62 &  1 & $20.68^{+ 0.15}_{- 0.09}$ & 0.31 & 0.39 & 0.54 \\
RECX-11 & 11616 & G. Herczeg & 0.0 & 10 & $19.70^{+ 0.03}_{- 0.20}$ & 0.03 & 0.04 & 0.06 \\
RECX-15 & 11616 & G. Herczeg & 0.0 & 10 & $19.58^{+ 0.05}_{- 0.18}$ & 0.02 & 0.03 & 0.04 \\
RU Lup (COS) & 12036 & J. Green & 0.07 & 11 & $20.05^{+ 0.03}_{- 0.06}$ & 0.07 & 0.09 & 0.13 \\
RU Lup (STIS - E140) & $8157-S$ & F. Walter & 0.07 & 11 & $20.23^{+ 0.10}_{- 0.33}$ & 0.11 & 0.14 & 0.19 \\
RW Aur A & 11616 & G. Herczeg & 1.58 & 7 & $20.25^{+ 0.05}_{- 0.21}$ & 0.11 &  0.15 & 0.20 \\
SU Aur & 11616 & G. Herczeg & 0.9 & 1 & $20.65^{+ 0.05}_{- 0.40}$ & 0.29 &  0.37 & 0.51 \\
SZ 102\tablenotemark{e} & 11616 & G. Herczeg & 1.13 & 12 & $20.58^{+ 0.05}_{- 0.18}$ & 0.24 & 0.31 & 0.43 \\
TWA 3A & 11616 & G. Herczeg & 0.7 & 13 & $19.20^{+ 0.10}_{- 0.20}$ & 0.01 & 0.01 & 0.02 \\
TW Hya & $8041-S$ & J. Linsky & 0.0 & 14 & $19.80^{+ 0.03}_{- 0.25}$ & 0.04 & 0.05 & 0.07 \\
UX Tau A & 11616 & G. Herczeg & 0.21 & 1 & $20.90^{+ 0.05}_{- 0.17}$ & 0.51 & 0.66 & 0.91 \\
V4046 Sgr & 11533 & J. Green & 0.0 & 15 & $19.85^{+ 0.03}_{- 0.28}$ & 0.04 & 0.06 & 0.08 \\
V836 Tau & 11616 & G. Herczeg & 1.68 & 7 & $20.70^{+ 0.28}_{- 0.46}$ & 0.32 & 0.42 & 0.57 \\

\enddata

\tablenotetext{a}{Program IDs with $-S$ are STIS observations.}
\tablenotetext{b}{~ (1) \citet{1995ApJS..101..117K}; (2) \citet{2006A&amp;A...459..837G}; (3) \citet{2003A&amp;A...409.1037A}; (4) \citet{2007ApJ...664L.111E}; (5) \citet{1992ApJ...385..217G}; (6) \citet{2003ApJ...583..334H}; (7) \citet{2004ApJ...616..998W}; (8) \citet{2000ApJ...544..927G}; (9) \citet{2007ApJ...665..512A}; (10) \citet{2004ApJ...609..917L}; (11) \citet{2005AJ....129.2777H}; (12) \citet{1994AJ....108.1071H}; (13) \citet{2001A&amp;A...369..239G}; (14) \citet{1983A&amp;A...121..217R}; (15) \citet{2000IAUS..200P..28Q}. }
\tablenotetext{c}{\cite{2006A&amp;A...459..837G} A$_{V}$(lit) values are calculated from $V$, $B-V$ magnitudes from \citet{1998A&amp;AS..133...81T}}
\tablenotetext{d}{This work, A$_{V}$(R$_{V}$) = (N(H I)/$4.8\times 10^{21}$) $\times$ R$_{V}$}
\tablenotetext{e}{Edge-on disk; emission is in reflected light}

\end{deluxetable*}

\begin{deluxetable}{cccc}\tablenum{3}
\tabletypesize{\scriptsize}
\tablecaption{Additional Model Parameters \label{tab:Model_Parameters}}
\tablewidth{0pt}
\tablehead{
\colhead{Object\tablenotemark{a} }& \colhead{FWHM$_{n}$} & \colhead{FWHM$_{b}$} & \colhead{$v_{HI}$}\\
&{km s$^{-1}$} &{km s$^{-1}$}&{km s$^{-1}$} }

\startdata
AA Tau & 560 & 610 & 10 \\
AB Aur & 650 & 1110 & 30 \\
AK Sco & 440 & 730 & -20 \\
BP Tau & 600 & 970 & 20 \\
CS Cha &  620 & 835 & 30 \\
CV Cha & 500 & 900 & 0 \\
DE Tau & 710 & 760 & -60 \\
DF Tau A & 740 & 965 & 20 \\
DK Tau A & 750 & 520 & -20 \\
DM Tau & 670 & 1020 & 30 \\
DN Tau & 550 & 1175 & 40 \\
DR Tau & 720 & 1160 & 40 \\
GM Aur & 340 & 730 & -60 \\
HD 100546 & 740 & 1310 & -50 \\
HD 104237 (E140M) &825 & 1260 & -40 \\
HD 104237 (G140M) & 570 & 850 & -40 \\
HD 135344B  & 535 & 1090 & -10 \\
HD 163296 & 580 & 1500 & 10 \\
HN Tau A & 450 & 830 & 40 \\
IP Tau &  510 & 760 & -70 \\
LkCa 15 & 650 & 795 & 50 \\
RECX-11 & 435 & 800 & 40 \\
RECX-15 & 500 & 800 & 50 \\
RU Lup (COS) & 510 & 750 & 50 \\
RU Lup (STIS - E140) & 550 & 950 & -30 \\
RW Aur A & 800 & 875 & 0 \\
SU Aur & 590 & 920 & -60 \\
SZ 102 & 700 & 760 & -50 \\
TWA 3A & 340 & 480 & -30 \\
TW Hya & 615 & 775 & -50 \\
UX Tau A & 550 & 800 & 10 \\
V4046 Sgr & 485 & 880 & -10 \\
V836 Tau & 550 & 750 & 0 \\

\enddata
\tablenotetext{a}{All targets have narrow-component heliocentric velocities of 49 km s$^{-1}$ and broad-component heliocentric velocities of 39 km s$^{-1}$ }

\end{deluxetable}
	
	In addition to these literature values, new optical extinctions are obtained here from fitting a combination of accretion continuum and weak-line T Tauri star (WTTS) photospheric templates to the observed optical emission.  The optical extinctions listed in Table 4 are obtained mostly from fits to broadband optical spectra \citep{HerczegHill2013} or, for earlier spectral types, to optical photometry and accurate spectral types \citep{2013MNRAS.429.1001A}.  Following \citet{HerczegHill2013}, optical extinctions for CV Cha, HD 104237, SU Aur, and AK Sco are recalculated here based on their spectral types and their flux-calibrated SITS spectra, while the extinction to RU Lup is calculated by assuming the blue continuum is flat.  The uncertainties in extinction are $\sim 0.2-0.3$ mag, which for most stars in our sample is attributable to uncertainty in spectral type.  The optical extinction estimates always assume a total-to-selective extinction of R$_{V}$ = 3.1.  Some variability in extinction has been documented for a few of our sources \citep{1994AJ....108.1906H, 1996AJ....112.2168S}, with AA Tau and RW Aur being a particularly notable examples \citep{2013arXiv1304.1487B, 2013AJ....146..112R}.  One source, SZ 102, is ignored in this analysis because the star is seen only through scattered light from the edge-on disk, which yields an unreliable extinction.  In section 4 we compare these dust derived estimates of A$_{V}$ to those derived above from the \ion{H}{1} column density. 

\begin{deluxetable*}{ccccccc}\tablenum{4}
\tabletypesize{\scriptsize}
\tablecaption{Additional A$_{V}$ Values \label{tab:OptFurXrayAvs}}
\tablewidth{0pt}
\tablehead{
\colhead{Object} & \colhead{A$_{V}$(3.1)} & \colhead{\citet{2011ApJS..195....3F}  A$_{V}$\tablenotemark{a}} & \colhead{Optical A$_{V}$} & \colhead{Optical Ref.\tablenotemark{b}} & \colhead{X-ray A$_{V}$\tablenotemark{c}} & \colhead{X-ray Ref.\tablenotemark{d}} \\
&{(This Work)}&&& }

\startdata
\multicolumn{7}{c}{}\\
AA Tau & 0.34 & 1.95 & 0.34 & 2 & 6.07 & 1 \\
AB Aur & 0.51 & 0.25\tablenotemark{d} & 0.53, 0.65 & 2,1 & 0.39 & 1 \\
AK Sco & 0.08 &  & 0.84 & 4 &  &  \\
BP Tau & 0.17 & 1.06 & 0.41& 2 & 0.58 & 1 \\
CS Cha & 0.16 &  &  & & 0.75 & 13 \\
CV Cha & 0.18 &  & 1.16 & 4 & 1.16 & 13 \\
DE Tau & 0.31 & 0.89 & 0.43 & 2 &  &  \\
DF Tau A & 0.54 & 1.95 & 0.12 & 2 &  &  \\
DK Tau A & 0.46 & 2.62 &  & & 2.00 & 1 \\
DM Tau & 0.48 & 0.0 & 0.09 & 2 & 0.78 & 1 \\
DN Tau & 0.18 & 0.92 & 0.52 & 2 & 0.39 & 1 \\
DR Tau & 0.48 & 1.42 &  & & 1.23 & 13 \\
GM Aur & 0.51 & 0.57 & 0.30 & 2 & 2.97 & 13 \\
HD 100546  & 0.09 &  &  &  & &  \\
HD 104237  & 0.08 &  & 0.01 & 4 & 0.50,1.1 & 2,3  \\
HD 135344B  & 0.12 &  & 0.96 & 1 & 0.55 & 13 \\
HD 163296 & 0.05 &  & 0.32 & 1 & 0.45,0.49 & 4,5 \\
HN Tau A & 0.36 & 1.06 &  & & 1.29 & 1 \\
IP Tau & 0.72 & 1.70 & 0.46 & 2 & 2.91 & 13 \\
LkCa 15 & 0.31 & 1.06 & 0.34 & 2 & 2.39 & 6 \\
RECX-11 & 0.03 &  & 0.0 & 3 & 0.04 & 7 \\
RECX-15 & 0.02 &  & 0.0 & 3 & 0.84 & 7 \\
RU Lup & 0.07 &  & 0.25 & 4 & 1.16 & 8 \\
RW Aur A & 0.11 & 0.50\tablenotemark{e} & & & 1.16  & 13 \\
SU Aur & 0.29 & 0.89\tablenotemark{f} & 0.67,0.76 & 2,4 & 2.13, 3.62 & 9,1 \\
SZ 102 & 0.24 &  &  & & &   \\
TWA 3A & 0.01 & & & & 0.06 & 10 \\
TW Hya & 0.04 &  & 0.0 & 3 & 0.13 & 11 \\
UX Tau A & 0.51 & 0.46 & 0.57 & 2 & 0.39 & 13 \\
V4046 Sgr & 0.04 &  &  & & 0.19 & 12 \\
V836 Tau & 0.32 & 1.49 & 0.64 & 2 & &   \\

\enddata
\tablenotetext{a}{Most values calculated from the \citet{2011ApJS..195....3F} A$_{J}$ values using the \citet{1985ApJ...288..618R} extinction law.}
\tablenotetext{b}{~ (1) \citet{2013MNRAS.429.1001A}; (2) \citet{HerczegHill2013}; (3) Default by association; (4) DAO STIS G430L data}
\tablenotetext{c}{Calculated from X-ray column densities using $N_{H}$ = (X-ray A$_{V}$) $\times$ ($4.8\times 10^{21}$/R$_{V}$) atoms cm$^{-2}$ mag$^{-1}$ assuming R$_{V} = 3.1$.  Some targets have multiple X-ray columns from the literature.}
\tablenotetext{d}{~ (1) \citet{2007A&amp;A...468..353G}; (2)\citet{2004ApJ...614..221S}; (3) \citet{2008ApJ...687..579T}; (4) \citet{2009A&amp;A...494.1041G}; (5) \citet{2005ApJ...628..811S}; (6) \citet{2013ApJ...765....3S}; (7) \cite{2010A&amp;A...524A..97L}; (8) \citet{Robrade_2007}; (9) \citet{2007A&amp;A...471..951F}; (10) \citet{2007ApJ...671..592H}; (11) \citet{2004A&amp;A...418..687S}; (12) \citet{2012ApJ...752..100A}; (13) This work.  See Section 3.4 and Table 5.}
\tablenotetext{d}{Calculated from the \citet{2003ApJ...590..357D} A$_{J}$ value cited in \citet{2011ApJS..195....3F} using the \citet{1985ApJ...288..618R} extinction law.}
\tablenotetext{e}{Calculated from the \citet{2001ApJ...556..265W} A$_{J}$ value cited in \citet{2011ApJS..195....3F} using the \citet{1985ApJ...288..618R} extinction law.}
\tablenotetext{f}{Calculated from the \citet{2004AJ....128.1294C} A$_{J}$ value cited in \citet{2011ApJS..195....3F} using the \citet{1985ApJ...288..618R} extinction law.}
\end{deluxetable*}

\subsection{X-ray Spectral Fitting}

	For the vast majority of young stars the typical available X-ray 
data consists of CCD resolution (R $\sim 15-20$)
spectra.  Such spectra are normally parameterized using 
global fitting tools, such as XSPEC \citep{1996ASPC..101...17A}. The 
X-ray emission from young stars is a combination of emission 
line and continuum emission. X-ray spectra for a 
range of young stars have been observed using grating 
spectroscopy with the {\it Chandra} HETG and LETG spectrographs 
and the {\it XMM-Newton} RGS spectrograph. These higher resolution spectra 
show the detailed emission line spectrum, which provides a firm 
foundation for modeling lower resolution CCD spectra, but usually 
provide only low signal-to-noise information on the continuum 
emission. Parameterization of CCD spectra can only be performed in 
limited ways, because the number of free parameters can quickly 
overwhelm the information content of the data. A common approach 
is to fit one or two temperature components and an interstellar 
column that introduces a low energy cutoff to the spectrum. 

	Two temperature components are almost always necessary for CCD 
spectra with a reasonable number of counts, because accreting 
T Tauri stars show two distinct sources of X-ray emission 
(originally discovered by \citealt{2002ApJ...567..434K}). A harder ($\sim$ 1 keV) 
component is produced 
within the hot coronal magnetic loops of a relatively standard 
active star corona, while a cooler ($\sim 0.2-0.3$ keV) 
component is associated with the highly-localized accretion  
shock hot-spot. The cooler component has a complex thermal structure 
whose properties are only crudely approximated by a single 
temperature.

	We have compiled all the available information of X-ray 
determined hydrogen column densities for 25 of the stars 
being studied in this paper (see Table 4), based on very similar 
two-temperature parameterization methods. Seventeen of these 
measurements are from the refereed literature, while for eight 
additional stars we 
present our new measurements in Table 5. Our  
two-temperature parameterizations were derived using XSPEC 
assuming sub-solar metalicities in the $0.2-0.3$ range.

\begin{deluxetable*}{ccccccccc}\tablenum{5}
\tabletypesize{\scriptsize}
\tablecaption{X-ray Measurements for Pre-Main Sequence Stars in our 
Sample\label{tab:newXrays}}
\tablewidth{0pt}
\tablehead{
\colhead{Star} & \colhead{Instrument} & \colhead{Obsid} &
\colhead{$N_{H}$} & \colhead{$kT_{1}$} & \colhead{$VEM_{1}$} &
 \colhead{$kT_{2}$} & \colhead{$VEM_{2}$}
&  \colhead{log L$_{X}$} \\
        &  &   &  \colhead{(10$^{21}$ cm$^{-2}$)} &  \colhead{(keV)} & 
  \colhead{(10$^{52}$ cm$^{-3}$)} &  \colhead{(keV)} & 
\colhead{(10$^{52}$ cm$^{-3}$)} & \colhead{(ergs s$^{-1}$)} }
\startdata
CS Cha   &Chandra& 6396 & 1.2$\pm$0.3 & 0.28$\pm$0.02 &  
47.5$\pm$20.8 & 1.03$\pm$0.05 & 26.3$\pm$3.4 & 30.41  \\
CV Cha   & XMM  & 0203810101 & 1.8$_{-0.4}^{+0.5}$ &0.8$\pm$0.1&
5.9$_{-1.2}^{+1.9}$& 3.2$_{-1.0}^{+3.0}$  & 8.3$\pm$1.5 &
30.3$\pm$0.1  \\
DR Tau   & XMM  & 0406570701 & 1.9$_{-0.5}^{+0.7}$ &0.8$\pm$0.2&
1.0$_{-0.3}^{+0.5}$& 3.5$_{-1.3}^{+4.4}$  & 1.9$_{-0.4}^{+0.5}$ &
29.6$\pm$0.1 \\
GM Aur   &Chandra & 8940 & 4.5$\pm$0.8 & 0.20$\pm$0.02 & 5.5$\pm$1.6 & 
0.98$\pm$0.06 & 11.5$\pm$1.6 & 29.86 \\
GM Aur   &Chandra & 9928 & 3.6$\pm$1.3 & 0.19$\pm$0.02 & 5.8$\pm$1.7 &
0.69$\pm$0.04 & 21.2$\pm$3.9 & 29.94 \\
GM Aur   &Chandra & 9929 & 4.6$\pm$2.1 & 0.30$\pm$0.04 & 12.3$\pm$5.4 &
1.73$\pm$0.26 & 5.4$\pm$1.2 & 29.67 \\
HD135344B&Chandra &  9927 & 0.85$\pm$0.48 & 0.12$\pm$0.04  &
1.4$\pm$4.9  & 0.63$\pm$0.03  & 3.9$\pm$1.3  & 29.44 \\
IP Tau   &Chandra & 10998 & 4.5$_{-1.5}^{+1.9}$& 0.18$\pm$0.03 &
99$_{-71}^{+473}$  & 1.0$_{-0.1}^{+0.2}$  & 7.1$_{-1.2}^{+1.7}$ &
30.1$_{-0.4}^{+0.6}$  \\
RW Aur A & XMM &0401870301& 1.9$\pm$1.0 &0.81$\pm$0.05& 4.3$\pm$0.7&
2.0$\pm$0.2  & 8.0$\pm$0.7 &
30.2$\pm$0.1 \\
UX Tau A & Chandra& 11001 & 0.6$_{-0.6}^{+0.9}$ &0.7$_{-0.2}^{+0.1}$&
6.2$_{-2.1}^{+2.8}$  & 1.6$_{-0.3}^{+0.8}$  & 8.5$\pm$2.1 &
30.3$_{-0.1}^{+0.2}$ \\
\enddata
\end{deluxetable*}
	
\begin{figure}\figurenum{5}
\begin{center}
\plotone{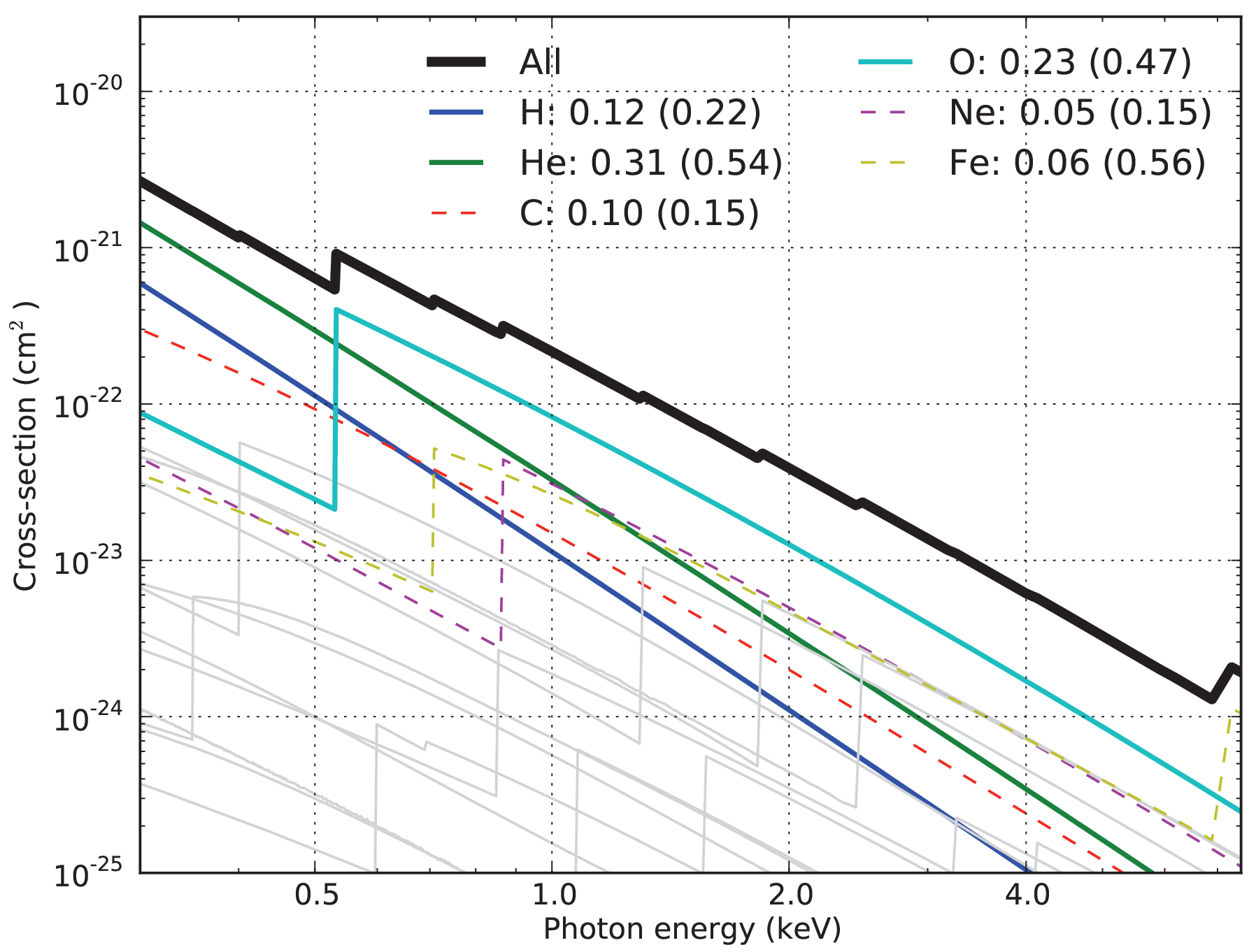}
\caption{Absorption cross-sections at X-ray wavelengths from \citet{1992ApJ...400..699B} with the contributions of the most important elements. Numbers give the mean fractional contributions to the total absorption cross-section in the displayed energy range (maximum fractions are given in parentheses). Thin gray lines indicate the absorption cross-sections of other elements. \label{fig:contributions}}
\end{center}
\end{figure}

	Compared to Lyman-$\alpha$ absorption, which directly traces the atomic hydrogen content, and optical/near-IR measurements, which trace the dust content, X-rays are attenuated by gas, grains, and molecules in the line of sight.  The absorption derived from X-ray observations is usually expressed as an equivalent hydrogen column density ($N_{H}$ = N(\ion{H}{1}) + 2N(H$_{2}$) + N(\ion{H}{2})), however the hydrogen column density is not directly measured.  Available X-ray observations of CTTS cannot distinguish individual elements of the absorber.  In fact, the equivalent $N_{H}$ is derived using an assumed (usually solar) abundance pattern and tabulated absorption cross-sections since hydrogen does not dominate the X-ray absorption in the energy windows of {\it Chandra} and XMM-Newton (0.3 -- 8\,keV or 2 -- 40\,\AA{}).  Therefore, the relation between measured X-ray absorption and column density does not involve empirical calibrations but depends only on the applied abundance pattern.  In the relevant energy range, the X-ray absorption cross-section is dominated by helium (up to 50\,\%) and oxygen (up to $\approx40\,$\%).  Depending on the considered wavelength, hydrogen provides only a small fraction ($\lesssim 22\,$\%) to the absorption cross-section (see Figure~\ref{fig:contributions}).  Because the hydrogen column density is not directly measured in the X-ray absorption, and only derived from the strong He and O absorption, it is difficult to distinguish an absorber with solar abundances of \ion{H}{1} from an absorber with fully ionized or fully molecular hydrogen.  We thus take the equivalent $N_{H}$ values in the X-ray to be our ``best estimate'' of N(\ion{H}{1}) on the line of sight to the region of X-ray production and compare these values to those calculated in our Lyman-$\alpha$ absorption modeling in Section 4.  	
	
\section{Results}

\begin{figure} \figurenum{6}
\begin{center}
\includegraphics[angle=90,scale=0.35]{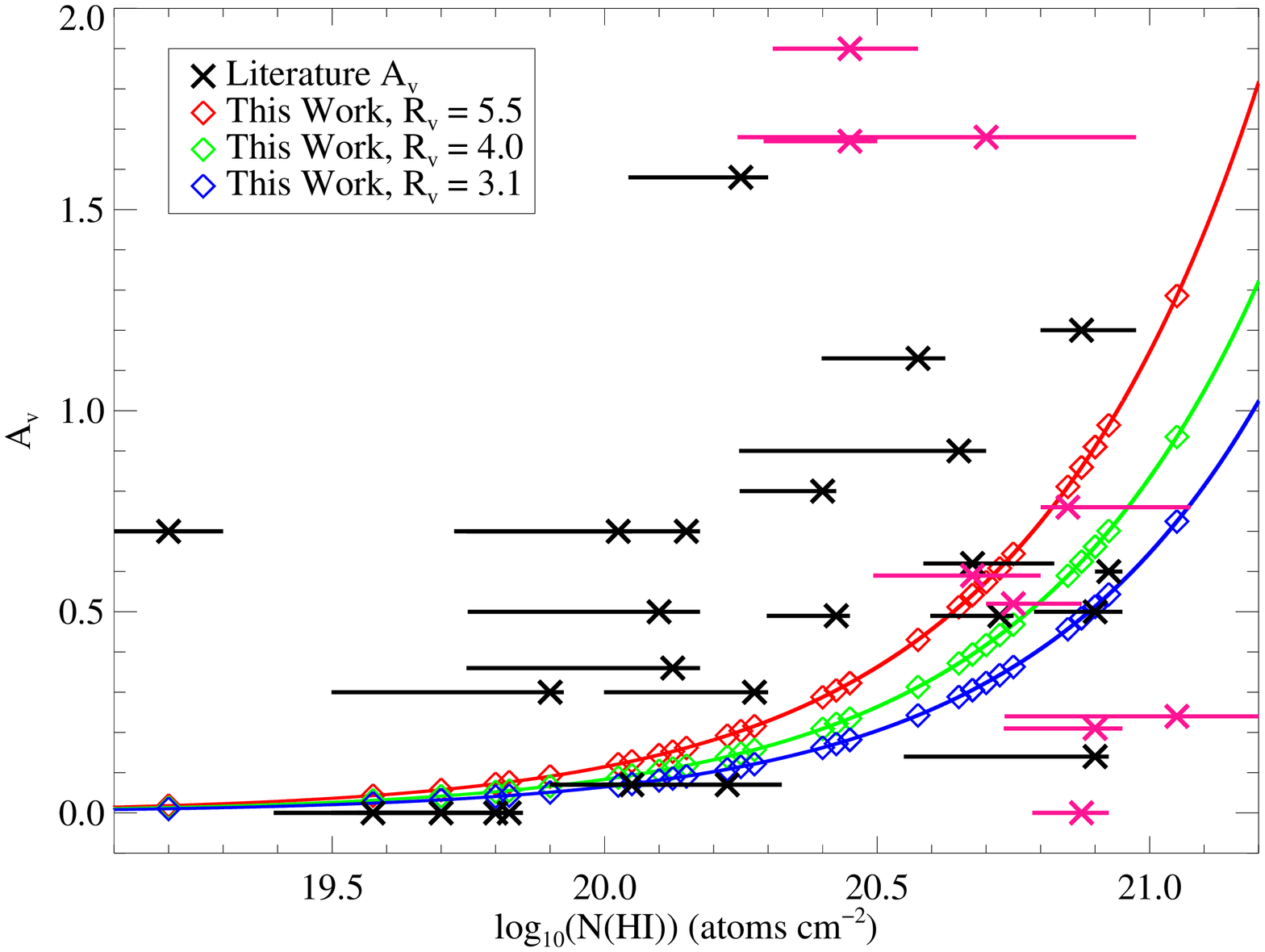}
\caption{
Literature values of A$_{V}$ compared against our computed \ion{H}{1} column densities (see Table 2). 
The black points are targets with reliable model fits, while the pink points (CV Cha, DE Tau, DK Tau, DM Tau, DN Tau, HN Tau, IP Tau, UX Tau, and V836 Tau) are targets with less reliable fits.  The \citet{1978ApJ...224..132B} relation between \ion{H}{1} column density and visual extinction for three different values of R$_{V}$ is overplotted with points corresponding to the A$_{V}$ values that are inferred in this work from the measured N(\ion{H}{1}).
The dust extinctions are typically higher than the gas extinctions.
}\label{fig:N(H)vsAv}
\end{center}
\end{figure}

	The \citet{1978ApJ...224..132B} relation as well as the relation between $E(B - V)$ and visual extinction, (R$_{V} =$ A$_{V}$/$E(B - V)$) yields a relation between column density and visual extinction:  N(\ion{H}{1})/A$_{V} = 4.8\times 10^{21}$/R$_{V}$ atoms cm$^{-2}$ mag$^{-1}$.  \citet{2004AJ....128.1294C} and \citet{2004ApJ...602..291W} find extinction laws for Taurus that are unlike the diffuse ISM (for which the \citet{1978ApJ...224..132B} relation is applicable).  Both papers find a weak to nonexistent 2175 \AA~ extinction bump.  However, in the far-UV, where the reddening correction is most critical, there is almost no difference between diffuse ISM reddening laws and those \citet{2004AJ....128.1294C} find more appropriate (see Figure 3 of \citealt{2004AJ....128.1294C}).  We include in Table 2 our calculated A$_{V}$ values from the N(\ion{H}{1}) - A$_{V}$ relation assuming R$_{V}$ values of 3.1, 4.0, and 5.5.  Errors are not assigned to the extinction values due to uncertainty in the choice of R$_{V}$.  We checked the Spearman's rank correlation coefficient and no correlation is seen between the N(\ion{H}{1}) values and the inclination of the targets (see the Appendix for a detailed description of inclinations).  This suggests that the majority of the \ion{H}{1} being measured along the line of sight toward our targets is likely interstellar.  We compare the literature A$_{V}$ values listed in Table 2 to our computed A$_{V}$ values in Figure~\ref{fig:N(H)vsAv}.\footnote{We compare our A$_{V}$ values using R$_{V}$ = 3.1 because all of the literature A$_{V}$ measurements assume that value of R$_{V}$.}  Targets with less reliable fits due to low signal-to-noise or negligible Lyman-$\alpha$ flux (CV Cha, DE Tau, DK Tau, DM Tau, DN Tau, HN Tau, IP Tau, UX Tau, and V836 Tau) are marked in pink in the plot.  

	Most of the extinction values in the literature are larger than the values calculated from our best-fit column densities.  For A$_{V}$ $\gtrsim 0.5$ mag, column densities N(\ion{H}{1}) $\gtrsim 10^{20.6}$ cm$^{-2}$ would be required.  The higher extinction values would require N(\ion{H}{1}) $\gtrsim 10^{21}$ cm$^{-2}$.  At these high column densities, attenuation along the line of sight would extinguish the stellar Lyman-$\alpha$ emission line completely \citep{2012ApJ...744...22F}, which may be happening in our highest column density target, IP Tau (N(\ion{H}{1}) = 10$^{21.05}$ cm$^{-2}$).  This fit is poor and may only be fitting continuum, in which case we could not reliably call this measurement an extinction value.  However, 19 out of the 26 CTTSs and all of the Herbig Ae/Be stars observed in our sample show unambiguous evidence for broad Lyman-$\alpha$ emission lines, arguing that the true interstellar \ion{H}{1} column densities to the Lyman-$\alpha$ emission must be lower than suggested by A$_{V}$ values in the literature.  Most of the 7 CTTSs with marginal Lyman-$\alpha$ emission detections (CV Cha, DK Tau, DN Tau, HN Tau, IP Tau, UX Tau, and V836 Tau) have large literature A$_{V}$ values.  However, as some of the extinction may be circumstellar, a non-interstellar dust-to-gas ratio in the disk (among other possibilities discussed in Section 5.2) may cause our column density measurements to be lower than expected from the published extinction values. 

\begin{figure} \figurenum{7}
\begin{center}
\includegraphics[angle=90,scale=0.35]{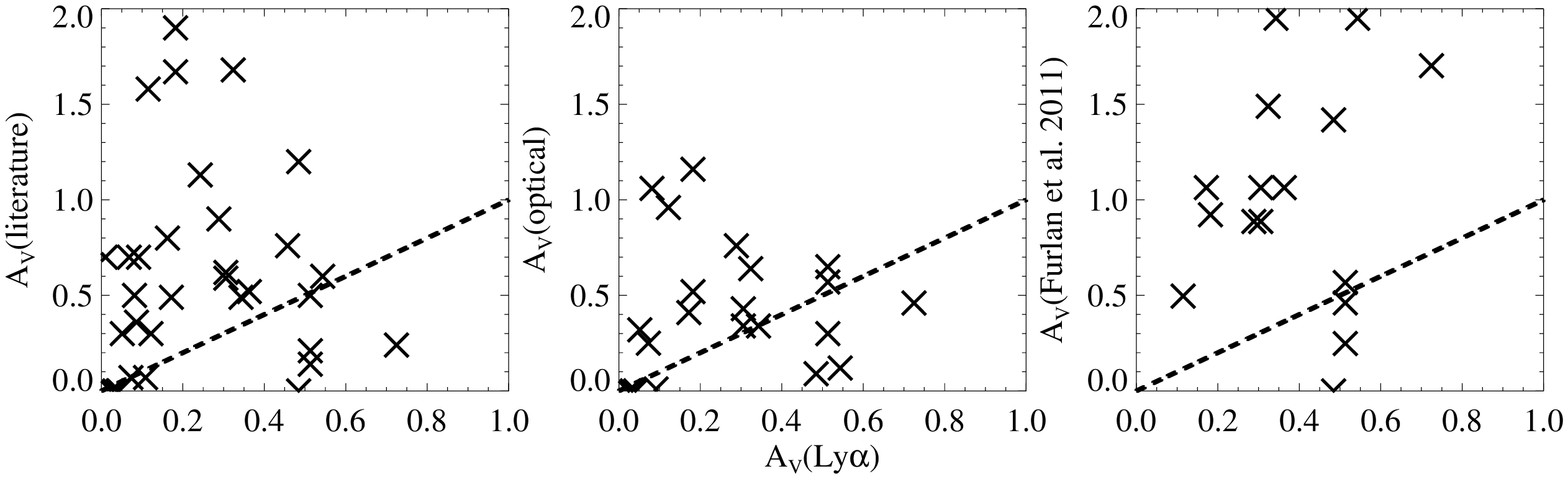}
\caption{
Comparison of our calculated visual extinction values from the far-UV analysis with the literature values from Table 2, as well as the optical, and IR \citep{2011ApJS..195....3F} values from Table 4.  Dashed lines showing a 1:1 correspondence are overplotted.  Our values are generally lower than any other visual extinction calculation.   
}\label{fig:Full_Extinction_Comparison}
\end{center}
\end{figure}

	In Figure~\ref{fig:Full_Extinction_Comparison}, we compare our calculated visual extinctions from the UV analysis (assuming R$_{V}$ = 3.1) to the literature values in Table 2 as well as infrared (IR) values from \citet{2011ApJS..195....3F} and optical values (both listed in Table 4 alongside our reprinted A$_{V}$ values for comparison).  The A$_{V}$ values from this work are generally lower than any other calculation of the visual extinction.  The optical extinctions are in best agreement with our values (though still larger for most targets), while the IR extinctions are consistently larger.  

		The reddening towards accreting K and M dwarf stars are difficult to accurately determine due to high amounts of veiling.  Changing the spectral type, accretion continuum shape, and the relative contribution of the two can greatly affect the A$_{V}$ value that is determined.  The early-type stars in our sample (AB Aur, AK Sco, CV Cha, HD 100546, HD 104237, HD 135344B, HD 163296, and SU Aur) should have minimal veiling affecting their optical spectra and relatively well-determined spectral types.  The large photospheric flux of the hot early-type stars dwarfs the flux from the accretion excess at optical wavelengths, while the cooler late-type stars generally have higher veiling \citep{1990ApJ...363..654B} due to their lower photospheric flux and consequently higher accretion flux to photospheric flux ratio.  This makes early-type stars critical to determine if the veiling is introducing large errors in the determination of the extinction.  We find that the discrepancy between the optical and Lyman-$\alpha$ determined extinction values is about the same in the early-type stars as the late-type stars in our sample.  We conclude that veiling alone cannot be the primary cause of the extinction discrepancies between optical and Lyman-$\alpha$ based measurements.   
		
\begin{figure} \figurenum{8}
\begin{center}
\includegraphics[angle=90,scale=0.35]{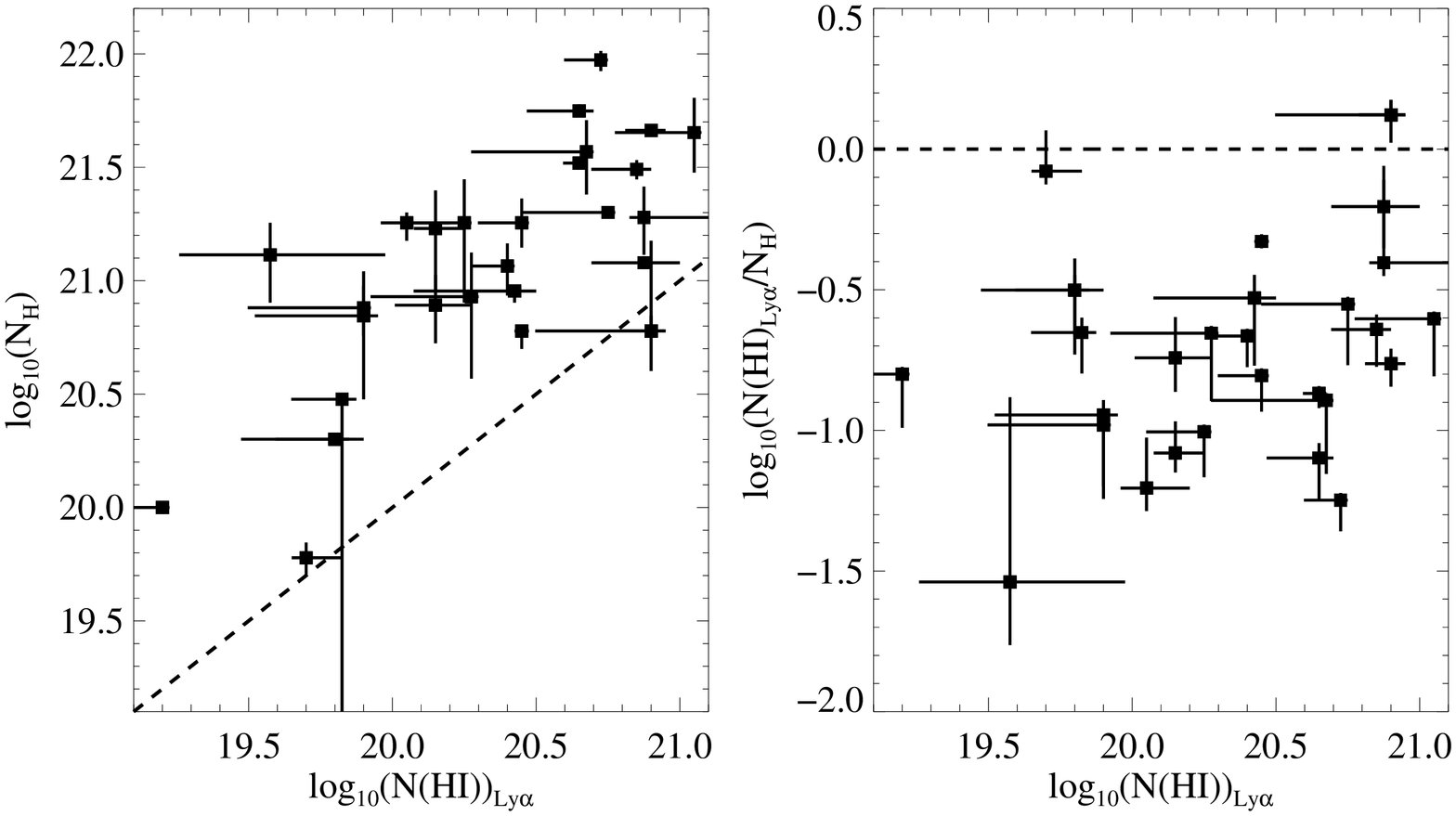}
\caption{
Comparison of X-ray equivalent hydrogen column densities ($N_{H}$) from the literature and our measurements with those from our Lyman-$\alpha$ absorption model fits.  The X-ray equivalent hydrogen column density is taken as our ``best estimate'' of N(\ion{H}{1}) derived from X-ray measurements.  A dashed line showing a 1:1 correspondence is overplotted.   
}\label{fig:Xray_Dual_Plot}
\end{center}
\end{figure}

	In Figure~\ref{fig:Xray_Dual_Plot} we compare the \ion{H}{1} column densities obtained through our Lyman-$\alpha$ fitting procedure to the equivalent hydrogen column density, $N_H$, obtained in the X-ray.  The X-ray columns are higher than the Lyman-$\alpha$ ISM columns in 23 out of 25 cases, implying that either a sizable fraction of circumstellar hydrogen does not contribute to the Lyman-$\alpha$ absorption (see the possibilities in Section 5.2), or that neutral hydrogen is depleted (most likely through ionization) with respect to O and He which dominate the X-ray absorption. Hydrogen is only of minor importance for the X-ray absorption so that changing the atomic hydrogen content has only a minor impact on the X-ray derived absorption.  The X-ray columns may also be higher due to the accretion shock punching deep into the dense photosphere of the star.  The soft X-rays produced in this shock region could thus probe the higher density photosphere, while the Lyman-$\alpha$ photons are emitted outside the densest shock region.  Much of the extra absorbing material is very hot and contains hydrogen that is mostly ionized so that the soft X-rays produce an equivalent hydrogen column density (see Section 3.4) that is an overestimation of the neutral hydrogen column density measured by the Lyman-$\alpha$ absorption.  We also detect in Figure~\ref{fig:Xray_Dual_Plot} an increase in the ratio between Lyman-$\alpha$ and X-ray columns with increasing Lyman-$\alpha$ column density.  This trend brings the two measurements closer to agreement as the Lyman-$\alpha$ column density increases, possibly suggesting an increased relative contribution to the total X-ray absorption from interstellar \ion{H}{1} on higher column density sightlines.        	
	
\section{Discussion}

\subsection{Sources of Hydrogen Along the Line of Sight}

\begin{figure} \figurenum{9}
\begin{center}
\plotone{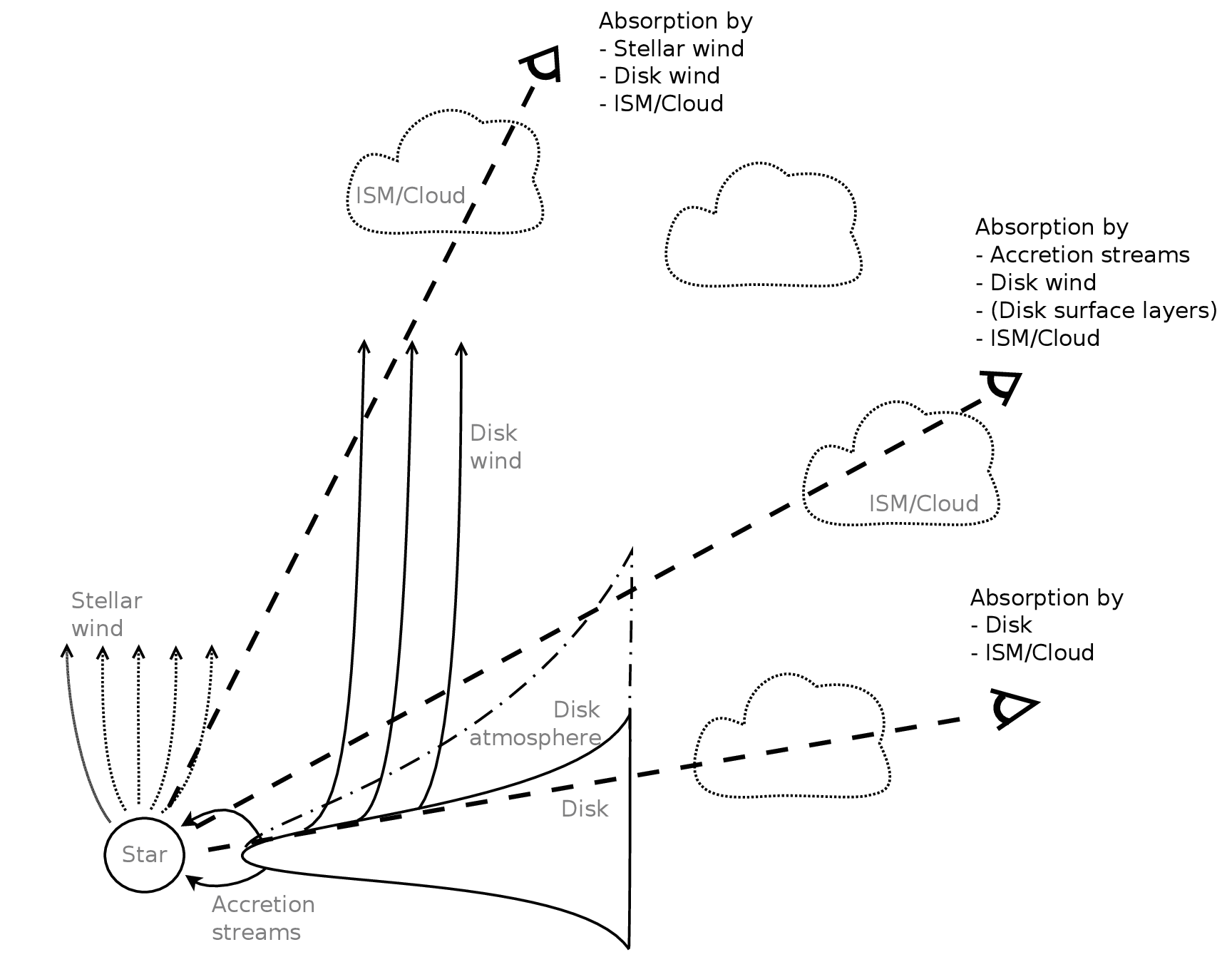}
\caption{
Sketch of the important components contributing to the absorption of the stellar emission.  The absorption depends strongly on the viewing geometry.  Lines with arrows indicate the plasma motion.  Sightlines are dashed.  
}\label{fig:Xray_Absorption}
\end{center}
\end{figure}

	Lyman-$\alpha$ absorption from partially ionized, primarily neutral, and primarily molecular gas can arise in several environments along the line of sight: in the protostellar outflow, in the disk atmosphere, and in the ISM.  Figure~\ref{fig:Xray_Absorption} illustrates the important components that may contribute to the absorption of the pre-main sequence stellar emission.   
A systematic error associated with our fits is the absence of an outflow component in our model.  Without a complete Lyman-$\alpha$ emission profile reconstruction (see e.g., \citealt{2004ApJ...607..369H, 2012ApJ...746...97S, 2012ApJ...756L..23S}), the Lyman-$\alpha$ outflow properties are challenging to determine.  An outflow absorbs mostly the blueward side of the Lyman-$\alpha$ emission (see Section 3.2), but can affect the redward side of the line in smaller amounts as well.  This leads to a systematic overestimation of the interstellar column density (and an underestimation of the total N(\ion{H}{1}), see Section 5.2) as our best-fit value is a combination of the true interstellar column density and \ion{H}{1} absorption from a protostellar outflow.	 

	\textit{Far-Ultraviolet Spectroscopic Explorer} ($FUSE$) measurements of H$_{2}$ along the line of sight to our Herbig Ae/Be stars (AB Aur, HD 100546, HD 104237, HD 135344B, HD 163296; \citealt{2008A&amp;A...484..225M}) can be combined with our \ion{H}{1} columns to check for agreement in the extinction values derived using the N(\ion{H}{1})-$E(B-V)$ and N(\ion{H}{1} + H$_{2}$)-$E(B-V)$ relations in \citet{1978ApJ...224..132B}.  They can also be combined to give a hydrogen column ($N_{H}$ = N(\ion{H}{1}) + 2N(H$_{2}$) + N(\ion{H}{2})) which directly compares with the equivalent hydrogen column measured by the X-ray absorption assuming a low ionization fraction.  The H$_{2}$ column densities measured in \citet{2008A&amp;A...484..225M} for our targets ($\sim 10^{16.5} - 10^{20.0}$ cm$^{-2}$) are generally much lower than our \ion{H}{1} columns, making $N_{H}$ only slightly larger than N(\ion{H}{1}).  So the $FUSE$ + $HST$ $N_{H}$ values are still lower ($\sim 0.8$ dex) than the X-ray $N_{H}$ values for all of the Herbig Ae/Be targets except AB Aur.  The AB Aur $FUSE$ value is 0.2 dex higher than the X-ray value due to its larger molecular hydrogen column density (log(N(H$_{2}$)) = 20.03; \citealt{2008A&amp;A...484..225M}).  Using the \citet{1978ApJ...224..132B} relation for a combined \ion{H}{1} and H$_{2}$ column, we obtained new extinction values for our Herbig Ae/Be targets that differed by 0.03 mag or less from those obtained from our \ion{H}{1} columns alone.  We conclude that the majority of this H$_{2}$ is most likely circumstellar rather than interstellar for our subset of targets based on their derived kinetic temperatures (56 - 758 K).  Only AB Aur has H$_{2}$ temperatures less than 300 K, suggesting that it is the only Herbig Ae/Be star in our sample with an unambiguous interstellar H$_{2}$ absorption component.  
		
	To determine the amount of \ion{H}{1} absorption that the circumstellar and interstellar material contribute along the line of sight, we compared N(\ion{H}{1}) measurements from hot, main-sequence stars, which should have little circumstellar material, to our Lyman-$\alpha$ measurements.  We hoped to use stars from \citet{1978ApJ...224..132B} and \citet{1994ApJS...93..211D} that sampled the Taurus-Auriga star-forming region ($168 < l < 181, -8 < b < -24, 80$ pc $< d < 220$ pc), but there were none to be found.  We instead searched only in distance, requiring the stellar distance to be between 80 pc and 220 pc, which led to a large sample of stars.  The average \ion{H}{1} column for the hot stars in the \citet{1978ApJ...224..132B} sample is log(N(\ion{H}{1})) = 20.27 and for those in the \citet{1994ApJS...93..211D} sample is log(N(\ion{H}{1})) = 20.92.  Our average \ion{H}{1} column is log(N(\ion{H}{1})) = 20.39.  The similarity in the average \ion{H}{1} columns in the three samples suggests that  most of the absorption we are measuring is interstellar rather than circumstellar.           
	
	Using our new extinction values calculated from \ion{H}{1} column densities, we can recalculate the intrinsic Lyman-$\alpha$ flux for the targets reported in \citet{2012ApJ...756..171F}.  Due to our extinction values generally being lower than those used by \citet{2012ApJ...756..171F} to reconstruct the Lyman-$\alpha$ flux, we find fluxes that can be up to $\sim 6500$ times smaller (DR Tau; \citet{2012ApJ...756..171F} assuming A$_{V}$ = 3.2 from \citealt{2009ApJ...704..531K}) than the Lyman-$\alpha$ fluxes reported in that work.  However, most target Lyman-$\alpha$ fluxes average between 2 and 300 times smaller with our new extinction values being used.  These lower flux values may greatly affect the chemistry of the gas in the protoplanetary disk region \citep{2009ApJ...690.1539G, 2009ApJ...705.1237G, 2012MNRAS.422.1880O, 2013arXiv1306.6623T}. 

\subsection{Discrepancy Among Lyman-$\alpha$, Optical, and IR-based extinction determinations}

	Several possibilities exist for the disagreement between our Lyman-$\alpha$ based extinction measurement (which samples the gas along the line of sight, through both the ISM and circumstellar material, to the Lyman-$\alpha$ emitting region) and the optical and IR-based extinctions (which measure the dust column to the stellar optical/IR photosphere).  
	
	{\it High dust-to-gas ratio -- } Some of the \ion{H}{1} absorption could be located in the stellar vicinity in material that has a large dust-to-gas ratio or a non-ISM extinction law.  In particular, the dust-to-gas ratio may be enhanced in a static disk through radial drift of dust grains \citep{2002ApJ...580..494Y}, and in star formation regions (e.g. \citealt{Vuong_2003}).  Because the dust-to-gas ratio may differ from that observed for the ISM in \citet{1978ApJ...224..132B}, the relation derived in that work may not be appropriate for the circumstellar environment.  \citet{2009ApJS..180..125R} fit direct measurements of H$_{2}$ column densities, \ion{H}{1} columns calculated from $E(B - V)$ values, and extinctions along translucent lines of sight, getting a relation very similar to \citet{1978ApJ...224..132B}.  However, our low \ion{H}{1} column densities may be in the same vicinity as large amounts of dust that has not settled (high dust-to-gas ratio) contributing to the high A$_{V}$ values measured in the optical and IR bands, making the \citet{1978ApJ...224..132B} relation provide inaccurate extinctions for our \ion{H}{1} columns.  
	
	{\it Geometric differences -- } It could also be possible that the Lyman-$\alpha$ emission is produced far out in the system (at large radii, $\gtrsim 100$ AU from the star).  The optical and IR observations may be probing the full extinction along the line of sight, whereas the Lyman-$\alpha$ profile that we observe the \ion{H}{1} absorption against may be generated further out in the system so that we are only probing the outer parts of the disk and ISM.  However, the large breadth of the Lyman-$\alpha$ profiles ($\Delta v$ $\gtrsim$ 500 km s$^{-1}$) points to an accretion origin for the emission, suggesting that the spatial differences between the two measurements may be only a few stellar radii.  RU Lup is an exception as the Lyman-$\alpha$ emission in the narrow (0.2" $\times$ 0.06") STIS slit was spatially extended and produced in the outflow.  Similarly, the broad Lyman-$\alpha$ emission (500 km s$^{-1}$) of HD 163296 in the STIS long-slit data extends out to a few 10's of AU, though is subdued compared to the central Lyman-$\alpha$ emission ($F_{broad}/F_{central} \sim 33 - 50$\%; P.C. Schneider (2013), private communication).   
	
	{\it Ionized or molecular hydrogen -- } The hydrogen may also be mostly ionized or molecular in the stellar vicinity.  If the hydrogen is mostly ionized so that there is little \ion{H}{1} absorption, there would need to be large quantities of dust in the same vicinity to contribute to the visual extinction.  This seems unlikely because any radiation that can ionize hydrogen should be strong enough to destroy dust grains as well.  If the hydrogen is ionized by a time-dependent outflow \citep{2007prpl.conf..277P}, however, the dust grains may not yet be destroyed.  The dust destruction may also be out of equilibrium \citep{1996ASPC..104..517F}, leading to large quantities of dust that is co-spatial with ionized hydrogen.  
	
	If the hydrogen is mostly molecular, there would need to be a large column of H$_{2}$ to provide the extra extinction to match the optical A$_{V}$ values listed in Table 4.  The average N(H$_{2}$)/N(\ion{H}{1}) ratio needed (using the \citet{1978ApJ...224..132B} relation) is N(H$_{2}$)/N(\ion{H}{1}) $\sim 1.7$, or a molecular fraction of 0.77, for the stars which have higher optical-based extinctions than Lyman-$\alpha$ derived extinctions.  
This molecular fraction is higher than any directly measured molecular fraction in the ISM measured in the UV, even on high-A$_{V}$ sightlines.  In particular, \citet{2002ApJ...577..221R, 2009ApJS..180..125R} and \citet{2007ApJ...658..446B, 2010ApJ...708..334B} find molecular fractions $\leq 0.76$ along translucent lines of sight.  However, a large molecular fraction may not be uncommon in a magnetized protostellar disk wind \citep{2012A&amp;A...538A...2P} and should be common in molecular clouds predicted by models (e.g. \citealt{1996ApJ...466..561M,2009A&amp;A...503..323V}), though no direct measurement yet exists.  
	 
	 Using N(H$_{2}$)/N(\ion{H}{1}) $\sim 1.7$ and our average derived \ion{H}{1} column density, the average required H$_{2}$ column density to match the optical extinctions for our targets is log(N(H$_{2}$)) $\sim 20.5$.  The T $\sim$ 2500 K H$_{2}$ absorber in the Lyman-$\alpha$ profile of AA Tau has log(N(H$_{2}$)) = 17.9 \citep{2012ApJ...744...22F} and the ubiquitous H$_{2}$ fluorescence observed by \citet{2012ApJ...756..171F} is explained by log(N(H$_{2}$)) $< 20.0$ as well.  The direct measurements of the Herbig stars in \citet{2008A&amp;A...484..225M} do find log(N(H$_{2}$)) $\gtrsim 20.0$ for a few targets that are not studied in this work (HD 141569, HD 176386, HD 259431, HD 38087, and HD 76543), but these all have modest molecular fractions ($\leq 0.35$).  However, the H$_{2}$ may be at T $\sim$ 500 K and cospatial with the CO absorption that is observed toward some CTTSs \citep{2013ApJ...766...12M}, which could possibly lead to a sufficient reservoir of H$_{2}$ to account for the additional reddening (assuming N(H$_{2}) \sim 10^{4}$ N(CO)).  
	
	If there is a large reservoir (log(N(H$_{2}$)) $\gtrsim 20.5$) of warm (T $\sim$ 500 K) molecular gas at a high molecular fraction ($f_{H2} \gtrsim 0.8$), then dust associated with this medium may be sufficient to explain the discrepancy between our Lyman-$\alpha$-based extinction measurements and those measured in the optical and IR.
	
	{\it Outflows -- } Lastly, smaller \ion{H}{1} column densities may be measured for an absorbing outflow that is not centered on our line of sight.  A geometry where the \ion{H}{1} is at the center of the Lyman-$\alpha$ absorption maximizes the \ion{H}{1} column density and could lead to a better agreement between the Lyman-$\alpha$ and optical/IR-based extinction values.  However, for most of our sources, the amount of \ion{H}{1} in the outflow is small compared to the amount of \ion{H}{1} in the ISM.  We can compare the ISM \ion{H}{1} column densities with the \ion{H}{1} outflow columns that were derived from the inner disk Lyman-$\alpha$ profile reconstructions presented by \citet{2012ApJ...756L..23S}.  For nearby sources like V4046 Sgr, N(\ion{H}{1})$_{ISM}$/N(\ion{H}{1})$_{outflow}$ $\sim 10$.  For a more distant source like DM Tau, N(\ion{H}{1})$_{ISM}$/N(\ion{H}{1})$_{outflow}$ $\sim 200$.  The average for all the targets in \citet{2012ApJ...756L..23S} is N(\ion{H}{1})$_{ISM}$/N(\ion{H}{1})$_{outflow}$ $\sim 74$.  However, because there is generally a larger optical depth on the blueward side of the Lyman-$\alpha$ line, the outflow N(\ion{H}{1}) may be significant for some sources.  
	
	An outflow could also have a high molecular fraction \citep{2012A&amp;A...538A...2P} or a high dust fraction, which could contribute to the extinction discrepancy as described above.  We calculated the amount of extinction that may be in the outflow by subtracting our Lyman-$\alpha$ \ion{H}{1} columns from the optical and \citet{2011ApJS..195....3F} extinctions.  We then used this difference to calculate an average gas-to-dust ratio in the outflow by comparing the extinction from the outflow to the outflow columns in \citet{2012ApJ...756L..23S}.  This gave us a gas-to-dust ratio for the outflow in the same manner as \citet{1978ApJ...224..132B}.  We find $\langle$N(\ion{H}{1})$_{outflow}/E(B - V)_{outflow} \rangle = 1.49 \times 10^{20}$ atoms cm$^{-2}$ mag $^{-1}$ (assuming R$_{V} = 3.1$), $\sim 30$ times lower than the \citet{1978ApJ...224..132B} gas-to-dust ratio, meaning the outflow would need to be very dusty in order to explain the large difference between our Lyman-$\alpha$ and the optical/IR-based extinctions.   
		
\subsection{Origin of X-ray Absorbing Gas}

	Emission in different spectral regions can originate 
from very different parts of the spatially complex young stellar
atmosphere and disk. The hot emitting gas seen in the FUV 
and X-ray regions can be produced in both the accretion 
hot-spot (which is thought to dominate in the FUV and 
soft-X-rays) and from more typical coronal magnetic loops (which 
produce most of the harder X-ray emission).  The absorbing column 
seen by different spectral features therefore depends on where 
they originate in the accretion shock. In the soft X-rays, 
systematically enhanced absorbing columns have been measured 
for emission lines produced within the concentrated accretion 
shock \citep{2010ApJ...710.1835B}.

	The absorption of the X-ray emission by the components shown in Figure~\ref{fig:Xray_Absorption} may lead to the excess X-ray column density that we detect.  One possibility for the observed excess X-ray absorption is that a sizable fraction of the hydrogen is ionized while helium or at least oxygen preserve a sufficiently low ionization, e.g., in the accretion streams connecting the inner edge of the disk and the stellar surface.  The inner parts of disks should have low ionization fractions  \citep{Dullemond_2001, Najita_2007}.  Disk winds should be mainly neutral initially \citep{1993ApJ...408..115S}, but temperatures behind internal shocks in a jet can be sufficient ($\sim 10^{6}$ K; \citealt{2007A&amp;A...468..515G, 2008A&amp;A...488L..13S}) to produce extreme ultraviolet and X-ray photons that significantly ionize the material coming off the disk surface.  Another possibility is that hydrogen is mainly molecular, located either in the outer parts of the protostellar system where the temperatures are sufficient for the existence of significant amounts of molecular hydrogen or in the ISM within the star forming region.  Lastly, some kind of hot stellar wind might be transparent to Lyman-$\alpha$ photons but not to X-ray photons.   	  
	
	The excess X-ray absorption has been studied previously in some of our targets.  Studies of the AA Tau system show that the absorption pattern of the circumstellar material differs from the ISM.  The AA Tau system is viewed at high inclination ($i\approx75^\circ$; \citealt{2007ApJ...659..705A}) and the star is periodically eclipsed by a disk warp.  The associated optical extinction can be directly measured \citep{Bouvier_2007} and the optical brightness variations are accompanied by periodic changes of the X-ray derived column density \citep{Schmitt_2007}.  However, the associated X-ray absorption is about ten times larger than expected based on A$_{V}$.  Due to the periodicity of the absorption pattern, it is possible to locate the associated absorber to a region about 0.1\,AU from the star, i.e., close to the dust sublimation radius for this system.  It is most likely that either the region around the disk warp is already dust-depleted or that the hot accretion streams provide the excess X-ray absorption since they are assumed to rigidly connect the star and the disk close to the radius of the disk warp.  Other key targets are TW~Hya and RU~Lup which are seen almost pole-on but also show excess X-ray absorption \citep{Robrade_2007, 2008A&amp;A...481..735G}.  \citet{Robrade_2007} interpret the extra X-ray absorption of RU~Lup as being related to accretion flows and an optically transparent wind emanating from the star or the disk while \citet{JK_2007} show that the FUV lines do not require a hot (stellar) wind in the case of TW~Hya.                     
	
\section{Conclusions}

	We present interstellar \ion{H}{1} column densities for 31 young stars determined from fitting absorption against the Lyman-$\alpha$ emission line.  We find that the literature A$_{V}$ values based on optical and IR observations are generally higher than the interstellar extinctions calculated from our derived column densities.  We also find that the Lyman-$\alpha$ derived column densities are smaller than the X-ray columns (which trace the gaseous part of the absorption like our Lyman-$\alpha$ measurements) for our targets.  Possible explanations for the extinction and column density discrepancies include 1) a high dust-to-gas ratio in the stellar vicinity, 2) the Lyman-$\alpha$ emission being produced far from the star, 3) the majority of the hydrogen being ionized or molecular, and 4) the N(\ion{H}{1}) absorption being dominated by a non-centered outflow.  The Lyman-$\alpha$ measurements determine the N(\ion{H}{1}) column density well, while the X-ray and optical extinctions may also be probing ionized and molecular gas along the line of sight.  Of these, we consider large dust-to-gas ratios or high molecular fractions to be the most likely.  However, these discrepancies may also arise because emission in the different spectral regions can be produced in many different regions around the star and disk.  Thus, these discrepancies do not necessarily imply that any measurement is wrong, but may be the result of the spatial complexity of the young star and disk system.  Our data suggests that the majority of the \ion{H}{1} absorption is interstellar in origin.  Targets with larger Lyman-$\alpha$ columns have better agreement between the Lyman-$\alpha$ and X-ray column density measurements.  Our lower visual extinctions lead to smaller Lyman-$\alpha$ fluxes in the protoplanetary region than previously calculated, possibly affecting the gas heating and chemistry in the disk.  Future work with an H$_{2}$ fluorescence model will enable us to constrain the shape of the far-UV extinction curve and possibly the grain-size distribution in these protoplanetary disks.  
	
\acknowledgments
	We thank Evelyne Roueff for helpful discussions pertinent to this work.  GJH appreciates a conversation with Sean Andrews and Sylvie Cabrit long ago on the reliability of inclination measurements.  This work made use of data from HST guest observing program 11616 and was supported by NASA grant NNX08AC146 to the University of Colorado at Boulder.

\appendix
\section{Appendix: Inclination Measurements}

\begin{figure} \figurenum{10}
\begin{center}
\includegraphics[angle=90,scale=0.6]{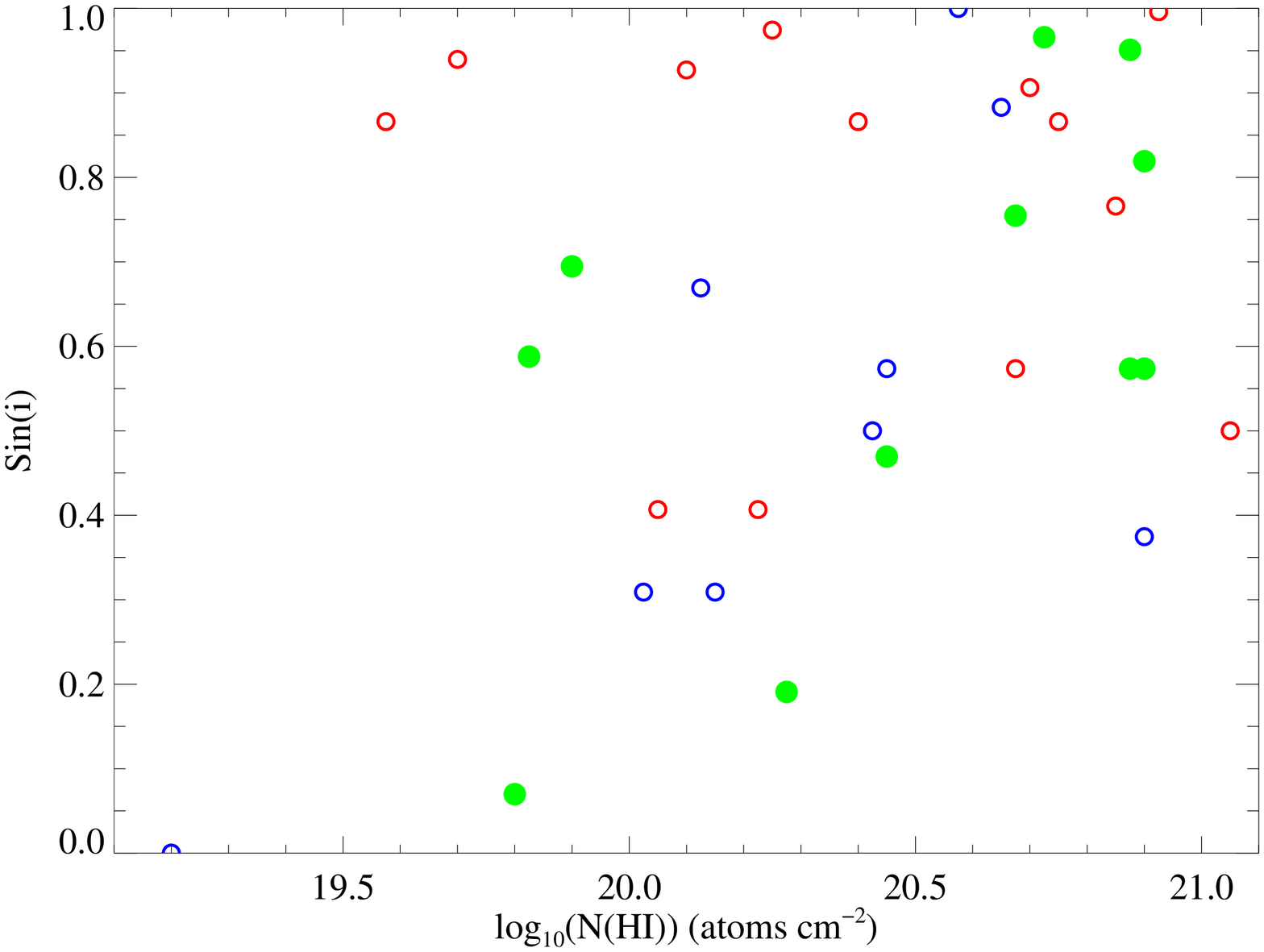}
\caption{
The sine of the literature disk inclinations listed in Table 6 (with 90$^{\circ}$ being edge-on) versus the logarithm of our measured \ion{H}{1} column densities.  Green filled circles are targets with the most accurate inclinations, red open circles are targets with less accurate inclinations, and blue open circles are the rest of the targets (other inclinations).     
}\label{fig:Inclinations}
\end{center}
\end{figure}

\begin{deluxetable*}{cccl}\tablenum{6}
\tabletypesize{\scriptsize}
\tablecaption{Inclination Methods \label{tab:IncMethods}}
\tablewidth{0pt}
\tablehead{
\colhead{Object} & \colhead{Inclination} & \colhead{Inclination Ref.\tablenotemark{a}} & \colhead{Inclination Method} \\
 & (degrees) & &  }

\startdata
AA Tau & 75\tablenotemark{b} & 3 & Elliptical Gaussian fits to the SMA sub-mm dust continuum visibilities \\
AB Aur &  22\tablenotemark{c} & 24 & Fit of IRAM 1.3mm continuum imaging \\
AK Sco & 68\tablenotemark{d} & 2 & Hipparcos parallax distance + orbit and stellar mass modeling (passive disk \\
 & & & models to SED)\\
BP Tau & 30\tablenotemark{c} & 22 & Chisquare minimization of models to IRAM interferometer observations of \\
 & & & $^{12}$CO J=2-1\\
CS Cha & 60\tablenotemark{d} & 8 & Assumption as no literature value was found\\
CV Cha & 35\tablenotemark{c} & 13 & Chisquare minimization of reconstructed brightness field maps of stellar surface \\
 & & & (spectropolarimetric observations)\\
DE Tau & 35\tablenotemark{d} & 14 & Spectroscopic IR stellar radii and rotation period + literature $v$sin$i$ in the red \\
DF Tau A & 85\tablenotemark{d} & 14 & Spectroscopic IR stellar radii and rotation period + literature $v$sin$i$ in the red \\
DK Tau A & 50\tablenotemark{d} & 14 & Spectroscopic IR stellar radii and rotation period + literature $v$sin$i$ in the red \\
DM Tau & 35\tablenotemark{b} & 4 & 2D Monte Carlo radiative transfer calculation leading to model of SMA \\
 & & & continuum visibilities \\
DN Tau & 28\tablenotemark{b} & 4 & 2D Monte Carlo radiative transfer calculation leading to model of SMA \\
 & & & continuum visibilities \\
DR Tau & 72\tablenotemark{b} & 3 & Elliptical Gaussian fits to the SMA sub-mm dust continuum visibilities \\
GM Aur & 55\tablenotemark{b} & 4 & 2D Monte Carlo radiative transfer calculation leading to model of SMA \\
 & & & continuum visibilities \\
HD 100546  & 42\tablenotemark{c} & 5 & Ellipticity of isophote fit to ACS coronographic images \\
HD 104237  & 18\tablenotemark{c} & 9 & Velocity (both in plane of sky and radial) of the knot A in the microjet \\
 & & & (STIS G140L) \\
HD 135344B  & 11\tablenotemark{b} & 16 & SMA CO spectral line image models ($^{12}$CO and $^{13}$CO J=2-1) \\
HD 163296 & 44\tablenotemark{b} & 21 & SMA and VLA mm-observation models of continuum, \\
 & & &  $^{12}$CO J=2-1 and J=3-2, and $^{13}$CO J=1-0 emission \\
HN Tau A & $>$ 60\tablenotemark{d} & 17 & Not well constrained.  Lower limit to intercept disk from flared disk model \\
IP Tau & 30\tablenotemark{d} & 6 & Stellar rotation velocity \\
LkCa 15 & 49\tablenotemark{b} & 4 & 2D Monte Carlo radiative transfer calculation leading to model of SMA \\
 & & & continuum visibilities \\
RECX-11 & 70\tablenotemark{d} & 15 & Magnetospheric accretion model fits to observed H-alpha profile  \\
RECX-15 & 60\tablenotemark{d} & 15 & Magnetospheric accretion model fits to observed H-alpha profile  \\
RU Lup & 24\tablenotemark{d} & 10,23 & Rotational broadening of absorption lines, Rotational period + R$_{star}$ + $v$sin$i$ \\
RW Aur A & 77\tablenotemark{d} & 7 & Interferometric data fit with inclined uniform disk model \\
SU Aur & 62\tablenotemark{c} & 1 & Visibility as a function of hour angle in the K-band fit with a Gaussian \\
 & & & brightness profile inclined on the sky \\ 
SZ 102 & $\sim 90$\tablenotemark{c} & 12 & Low luminosity possibly due to blocking of direct light by edge-on disk \\
TWA 3A & $\sim 0$\tablenotemark{c} & 11 & Assumption from large mid-IR excess and negligible optical reddening \\
TW Hya & 4 - 10\tablenotemark{b} & 19 & Disk models of ALMA $^{12}$CO J=2-1 and J=3-2 data \\
UX Tau A & 35\tablenotemark{b} & 4 & Fit of bright, resolved CO J=3-2 disk in SMA data \\
V4046 Sgr & 33.5\tablenotemark{b} & 20 & Models of SMA $^{12}$CO J=2-1 data \\
V836 Tau & 65\tablenotemark{d} & 18 & Assumed from colors and errors on rotation period, R$_{star}$, and $v$sin$i$ \\

\enddata
\tablenotetext{a}{~ (1) \citet{2002ApJ...566.1124A}; (2) \citet{2003A&amp;A...409.1037A}; (3) \citet{2007ApJ...659..705A}; (4) \citet{2011ApJ...732...42A}; (5) \cite{2007ApJ...665..512A}; (6) \citet{2013arXiv1304.3746A} derived from \citet{2005ESASP.560..571G}; (7) \citet{2007ApJ...669.1072E}; (8) \citet{2007ApJ...664L.111E, 2011ApJ...728...49E}; (9) \citet{2004ApJ...608..809G}; (10) \citet{2005AJ....129.2777H}; (11) \citet{2007ApJ...671..592H}; (12) \citet{1994AJ....108.1071H}; (13) \citet{2009MNRAS.398..189H}; (14) \citet{2001ApJ...561.1060J}; (15) \citet{2004MNRAS.351L..39L}; (16) \citet{2011AJ....142..151L}; (17) \citet{2013ApJ...766...12M}; (18) \citet{2008ApJ...687.1168N}; (19) \citet{2012ApJ...757..129R}; (20) \citet{2012ApJ...759..119R}; (21) \citet{2007A&amp;A...469..213I}; (22) \citet{2000ApJ...545.1034S}; (23) \citet{2007A&amp;A...461..253S}; (24) \citet{2012A&amp;A...547A..84T}.}
\tablenotetext{b}{~Most accurate inclination}
\tablenotetext{c}{~Other inclination}
\tablenotetext{d}{~Less accurate inclination}

\end{deluxetable*}

	In Figure~\ref{fig:Inclinations}, we compare the sine of the literature inclination values listed in Table 6 (with 90$^{\circ}$ being edge-on) to the logarithm of our Lyman-$\alpha$ based \ion{H}{1} column densities.  A number of our disks are transitional or pre-transitional, which may have mm-sized grains that are not azimuthally-symmetric in their distribution \citep{2013Natur.493..191C, 2013arXiv1309.7400F, 2013Sci...340.1199V}, making dust-based inclination estimates less reliable.  Also, many targets have both gas and dust disks, which may provide different inclinations.  However, for the transitional and pre-transitional disks in our sample with purely dust-based inclinations quoted in Table 6 (DM Tau, GM Aur, and LkCa 15), we find gas-based inclinations in the literature that differ by only a few degrees from those determined by the dust (DM Tau: 33$^{\circ}$, \citealt{1998A&amp;A...339..467G}; GM Aur: 56$^{\circ}$, \citealt{2000ApJ...545.1034S}; LkCa 15: 52$^{\circ}$, \citealt{2000ApJ...545.1034S}).  We adopt the inclination values in Table 6 for these targets, although there is little difference.  The other transitional and pre-transitional disks with the most accurate inclinations have gas-based or both gas and dust-based inclinations.  For the targets with the most accurate inclination values (see below), there is no tight correlation (based on the Spearman's rank correlation coefficient) with the \ion{H}{1} column densities.  Since we would expect a correlation to exist if \ion{H}{1} absorption was coming from the disk (see Figure~\ref{fig:Xray_Absorption}), this suggests that the majority of our measured \ion{H}{1} column density is interstellar.         
		
\subsection{Most accurate Inclination Measurements}

	The most accurate inclination measurements come from spatially and spectrally resolved CO emission lines in the sub-mm \citep{2011AJ....142..151L, 2012ApJ...759..119R, 2012ApJ...757..129R} and spatially resolved disks in sub-mm dust continua \citep{2007ApJ...659..705A, 2007A&amp;A...469..213I, 2011ApJ...732...42A}.  2D model fits to the data provide strong constraints on the orientation of the disk and are a direct method of determining the inclinations.  Out of our 31 targets, 11 have the most accurate inclinations. 

\subsection{Less accurate Inclination Measurements}

	Some of the less accurate inclinations found in the literature were derived from stellar rotation data.  These measurements rely on accurate measurements of $v$sin$i$, the stellar rotation period, and the stellar radius.  The rotation period and the value of $v$sin$i$ can be measured much more accurately than the stellar radius, which depends on the often uncertain parameters of stellar luminosity, reddening, veiling, and effective temperature (see Section 1).  \citet{2013A&amp;A...558A..83A} find that inclination measurements determined from the stellar rotation are less reliable than those determined from spatially resolved disk observations.  
	
	Other less accurate inclinations come from SED modeling (AK Sco; \citealt{2003A&amp;A...409.1037A}), an assumption as no literature values were found (CS Cha; \citealt{2007ApJ...664L.111E, 2011ApJ...728...49E}), lower limits to intercept a flared disk model (HN Tau; \citealt{2013ApJ...766...12M}), H$\alpha$ profile fits (RECX-11, RECX-15; \citealt{2004MNRAS.351L..39L}), and fits to interferometric data (RW Aur; \citealt{2007ApJ...669.1072E}).  Out of our 31 targets, 12 have less accurate inclinations.  
	
\subsection{Other Inclination Measurements}

	The remaining inclination measurements do not spatially resolve the disk itself, but may be more accurate than the less accurate inclinations.  These include targets such as AB Aur which has an inclination angle that varies with scale \citep{2012A&amp;A...547A..84T}, CV Cha which has an inclination measured from $\chi^{2}$ minimization models fit to brightness field maps \citep{2009MNRAS.398..189H}, and HD 104237 which has the inclination constrained using a microjet in the system \citep{2004ApJ...608..809G}.  Other inclinations are those for BP Tau, HD 100546, SU Aur, SZ 102, and TWA 3A.  Though SZ 102 and TWA 3A are typically quoted in the literature as edge-on and face-on, respectively, a well-defined measurement has not yet been made to our knowledge.  Out of our 31 targets, 8 fall in this inclination category.

\bibliography{Av_paper}

\end{document}